\documentclass[useAMS,usenatbib]{mn2e}
\usepackage{epsf}
\usepackage{graphicx}

\newcommand{\equ}[1]{eq.~(\ref{eq:#1})}
\newcommand{\equs}[1]{eqs.~(\ref{eq:#1})}
\newcommand{\equm}[1]{(\ref{eq:#1})}

\newcommand{\se}[1]{\S\ref{sec:#1}}
\newcommand{\fig}[1]{Fig.~\ref{fig:#1}}

\newcommand{\Fig}[1]{Figure~\ref{fig:#1}}
\newcommand{\Figs}[1]{Figures~\ref{fig:#1}}
\newcommand{\be}{\begin{equation}}
\newcommand{\ee}{\end{equation}}
\newcommand{\bea}{\begin{eqnarray}}
\newcommand{\eea}{\end{eqnarray}}

\newcommand{\msun}{M_\odot}
\newcommand{\Msun}{M_\odot}

\newcommand{\ifm}[1]{\relax\ifmmode#1\else$\mathsurround=0pt #1$\fi}
\newcommand{\kms}{\ifmmode\,{\rm km}\,{\rm s}^{-1}\else km$\,$s$^{-1}$\fi}
\newcommand{\hmpc}{\,\ifm{h^{-1}}{\rm Mpc}}
\newcommand{\hkpc}{\,\ifm{h^{-1}}{\rm kpc}}
\newcommand{\Mpc}{\,{\rm Mpc}}
\newcommand{\kpc}{\,{\rm kpc}}
\newcommand{\pc}{\,{\rm pc}}

\newcommand{\ltsima}{$\; \buildrel < \over \sim \;$}
\newcommand{\lsim}{\lower.5ex\hbox{\ltsima}}
\newcommand{\gtsima}{$\; \buildrel > \over \sim \;$}
\newcommand{\gsim}{\lower.5ex\hbox{\gtsima}}
\newcommand{\prop}{\propto}
\newcommand{\dd}{\rm d}

\newcommand{\rar}{\rightarrow}

\def\Mv{M_{\rm v}}
\def\Rv{R_{\rm v}}
\def\Vv{V_{\rm v}}
\def\rhov{\rho_{\rm v}}
\def\omm{\Omega_{\rm m}}
\def\oml{\Omega_{\Lambda}}

\def\sy{\,M_\odot\, {\rm yr}^{-1}}
\def\cmmc{\,{\rm cm}^{-3}}
\def\cmpc{\,{\rm cm}^{3}}
\def\cmms{\,{\rm cm}^{-2}}
\def\lya{${\rm L}\alpha$\ }
\def\La{{\rm L}\alpha}
\def\ergs{\,{\rm erg\, s}^{-1}}

\def\ergsc{\,{\rm erg\, s}^{-1}\,{\rm cm}^{-2}\,{\rm arcsec}^{-2}}
\def\fphi{\hat\phi}

\def\fc{f_{\rm c}}
\def\fg{f_{\rm grav}}
\def\fa{f_\alpha}

\def\II{\rm {II}}
\def\H{\rm {H}}
\def\HI{\rm {HI}}
\def\HII{\rm {HII}}

\def\HeII{\rm {HeII}}
\def\HeIII{\rm {HeIIII}}

\def\MN{MN\,}
\def\CD{CDB\,}

\def\taudust{\tau_{\rm dust}}

\def\falph{0.85}
\def\fcfa{0.34}

\usepackage{color}

\title[Cold streams as Lyman-alpha blobs]
{Gravity-driven Lyman-alpha blobs from cold streams into galaxies}

\author[Goerdt, Dekel, Sternberg, Ceverino, Teyssier \& Primack]
{\parbox[t]{\textwidth}{Tobias Goerdt$^1$\thanks{tgoerdt@phys.huji.ac.il},
A. Dekel$^1$\thanks{dekel@phys.huji.ac.il}, 
A. Sternberg$^2$\thanks{amiel@wise.tau.ac.il}, 
D. Ceverino$^1$\thanks{ceverino@phys.huji.ac.il},
R. Teyssier$^3$\thanks{teyssier@physik.uzh.ch} \& \\
J. R. Primack$^4$\thanks{joel@scipp.ucsc.edu}} \\ \vspace*{3pt} \\
$^1$Racah Institute of Physics, The Hebrew University, Jerusalem 91904,
Israel \\
$^2$The Raymond and Beverly Sackler School of Physics and Astronomy, 
Tel Aviv University, Tel Aviv 69978,
Israel \\
$^3$Institut f\"ur Theoretische Physik, Universit\"at Z\"urich, Winterthurer
Strasse 190, CH-8057 Z\"urich, Schweiz \\
$^4$Department of Physics, University of California, Santa Cruz, CA 95064, USA
}

\date{Draft version \today}
\pagerange{\pageref{firstpage}--\pageref{lastpage}}
\pubyear{2009}

\begin{document}

\maketitle

\label{firstpage}

\begin{abstract}
We use high-resolution cosmological hydrodynamical AMR simulations to predict
the characteristics of \lya emission from the cold gas streams that fed
galaxies in massive haloes at high redshift. The \lya luminosity in our
simulations is powered by the release of gravitational energy as gas flows from
the intergalactic medium into the halo potential wells. The UV background
contributes only $<20\%$ to the gas heating. The \lya emissivity is due
primarily to electron-impact excitation cooling radiation in gas $\sim 2\times
10^4$~K. We calculate the \lya emissivities assuming collisional ionisation
equilibrium (CIE) at all gas temperatures. The simulated streams are
self-shielded against the UV background, so photoionisation and recombination
contribute negligibly to the \lya line formation. We produce theoretical maps
of the \lya surface brightnesses, assuming that $\sim 85\%$ of the \lya photons
are directly observable. We do not consider transfer of the \lya radiation, nor
do we include the possible effects of internal sources of photoionisation such
as star-forming regions. Dust absorption is expected to obscure a small
fraction of the luminosity in the streams. We find that typical haloes of mass
$\Mv\sim10^{12-13}\msun$ at $z\sim 3$ emit as \lya blobs (LABs) with
luminosities $10^{43-44}\ergs$. Most of the \lya comes from the extended
($50-100\kpc$) narrow, partly clumpy, inflowing, cold streams of $(1-5)\times
10^4$K that feed the growing galaxies. The predicted LAB morphology is
therefore irregular, with dense clumps and elongated extensions. The integrated
area contained within surface-brightness isophotes of $2 \times 10^{-18}\ergsc$
is $\sim 2-100\, {\rm arcsec}^2$, consistent with observations. The linewidth
is expected to range from $10^2$ to more than $10^3 \kms$ with a large
variance. The typical \lya surface brightness profile is $\propto r^{-1.2}$
where $r$ is the distance from the halo centre. Our simulated LABs are similar
in luminosity, morphology and extent to the observed LABs, with distinct
kinematic features. The predicted \lya luminosity function is consistent with
observations, and the predicted areas and linewidths roughly recover the
observed scaling relations. This mechanism for producing LABs appears
inevitable in many high-$z$ galaxies, though it may work in parallel with other
mechanisms. Some of the LABs may thus be regarded as direct detections of the
cold streams that drove galaxy evolution at high $z$.
\end{abstract}


\section{Introduction}
\label{sec:intro}
Hundreds of Lyman-alpha blobs (LAB) have been detected so far in the redshift
range $z = 2 - 6.5$, mostly near $z \sim 3$ \citep{steidel, matsuda, saito,
ouchi, yang}. Their luminosities range from below $10^{43}$ to above
$10^{44}\ergs$, and they extend on the sky to $30-50\kpc$ and more. The two
main open questions are (a) the origin of the extended cold and relatively
dense gas capable of emitting \lya, and (b) the continuous energy source for
exciting the gas to emit \lya.

The emitting hydrogen gas should be at a temperature $T \gsim 10^4$K,
relatively dense and span a much larger area than covered by the stellar
component of galaxies. It may arise from outflows or inflows. The energy
sources discussed in the literature include photoionisation by obscured AGNs,
early starbursts or extended X-ray emission \citep{haimanb, jimenez, scharf},
as well as compression of ambient gas by superwinds to dense \lya emitting
shells \citep{mori}, and star formation triggered by relativistic jets from
AGNs \citep{rees}.
  
Many of the bright LABs are found in the vicinity of massive, star-forming
galaxies \citep{matsudab}. Multi-wavelength observations reveal that a fraction
of the LABs are associated with sub-millimetre and infrared sources that
indicate very high star-formation rates (SFR) in the range $10^2-10^3 \sy$ 
\citep{chapman, geacha, geachb} or with obscured active galactic nuclei (AGN) 
\citep{basu, scarlata}. While stellar feedback and AGNs could in principle
provide the energy source for the \lya luminosity, many LABs are not associated
with sources of this sort that are powerful enough to explain the observed \lya
luminosities \citep{kim, sj}. This indicates that star formation and AGNs are
not the sole drivers of LABs, and may not even be the dominant source for LABs.

Indeed, high-redshift galaxies exhibit a generic mechanism that simultaneously
provides both the cold gas and the energy source. It is a direct result of the
phenomenon robustly established by simulations and theoretical analysis, where
high-z massive galaxies are continuously fed by narrow, cold, intense, partly
clumpy, gaseous streams that penetrate through the shock-heated halo gas into
the inner galaxy, grow a dense, unstable, turbulent disc with a bulge and 
trigger rapid star formation \citep{bd03, keresa, db06, ocvirk, keresb, nature,
peter, dsc, cd}. Massive clumpy star-forming disks observed at $z\sim 2$
\citep{genzel08, genel, foerster2} may have been formed via the smooth and
steady accretion provided by cold flows, as opposed to merger events. The
streaming of the gas into the dark-matter halo potential well is associated
with transfer of gravitational energy to excitations of the hydrogen atoms
followed by cooling emission of \lya \citep{haiman, fardal, db06, db08, ko08,
furlanetto2, furlanetto}. 

There were two earlier attempts to compute the \lya cooling radiation from
hydrodynamic SPH cosmological simulations. Based on their analysis,
\citet{fardal} pointed out the potential association of this \lya cooling 
emission with the first observed LABs. These early simulations did not allow a
proper resolution of the detailed structure of the cold streams as they
penetrate through the hot medium. Their shortcomings included intrinsic
limitations of the SPH technique, a limited force resolution of $7 \hkpc$
(comoving), not allowing for radiative cooling below $10^4$K and neglecting the
photoionisation by the UV background. \citet{furlanetto} used SPH simulations
with a somewhat higher resolution of $\sim$1\,kpc to make more detailed
comparisons of the simulated and observed luminosity functions and size
distributions of LABs powered by cold accretion. In their pessimistic model,
assuming no emissivity from gas that is self-shielded from the ionising
background radiation, they end up with low \lya luminosity. They comment that
cooling IGM gas may explain the observations only if one adopts an optimistic
scenario, where the self-shielded gas is emitting \lya at CIE. The latter
scenario will also be adopted in this paper. Other sources of ionisation are
ignored and radiative transfer is not applied.

\citet[][hereafter DL09]{dijkstra} have recently worked out an analytic toy
model for \lya cooling radiation from cold streams, based on the general
properties of the cold streams as reported from cosmological simulations. They
conclude that the streams could in principle provide spatially extended \lya
sources with luminosities, line widths and abundances that are similar to those
of observed LABs. They point out that the filamentary structure of cold flows
may explain the wide range of observed LAB morphologies. They also highlight
the fact that the most luminous cold flows are associated with massive haloes,
which preferentially reside in dense large-scale surroundings, in agreement
with the observed presence of bright LABs in dense environments. This model
presents a successful feasibility test for the role of cold streams in powering
the LAB emission, and it provides physical intuition into the way by which this
process could be manifested. However, a comparison to our simulations indicates
that this simplified model does not capture the detailed hydrodynamic
properties of the cold streams. In particular, it significantly over-predicts
the cold-gas density, under-predicts the gas temperature, and does not account
for the partly clumpy nature of the streams and their characteristic radial
distribution in the halo. As a result, the DL09 model underestimates the total
\lya luminosity by a factor of a few, and it therefore has to appeal to the
excessive clustering of the LABs in order to boost up the predicted luminosity
function for a match with the observations.

In the current paper, we calculate the \lya emission from the cold gas in 
high-redshift galactic haloes using state-of-the-art hydrodynamical AMR
cosmological simulations. With a maximum resolution better than 70\,pc in our
simulations we can quite accurately map the extended \lya sources. This allows
us to study their individual shapes, morphologies and kinematics. We measure
quantities such as the distribution of surface brightness within each halo, the
area covered with surface brightness above a given isophotal threshold, the
predicted observed linewidth, the typical total \lya luminosity per halo of a
given mass, and the overall \lya luminosity function of LABs. These predicted
properties are compared with the observed LABs. 

This paper is organised as follows. In \se{feasibility} we work out a simple
feasibility test where we use a simple toy model to estimate the expected
gravitational heating power as a proxy for the \lya luminosity. In \se{sim} we
introduce the two sets of cosmological simulations used. In \se{lal} we explain
our methodology of computing \lya emissivity and luminosity as a function of
halo mass. In \se{lyagas} we identify the gas that contributes to the \lya
emission. In \se{images} we apply our methodology to the simulations and
provide predicted images of LABs. In \se{L(M,z)} we determine the scaling of
\lya luminosity as a function of halo mass and redshift. In \se{lumfun} we
compare our predicted \lya luminosity function with observational results. In
\se{obs} we measure the predicted surface-density profile, isophotal area and
linewidth and compare to observations. In \se{grav} we show that gravitational
heating is the main source of energy driving the \lya luminosity in our
simulations, while the role of photoionisation is minor. In \se{con} we discuss
our analysis and results and draw our conclusions.

\section{Feasibility of Gravitational Heating}
\label{sec:feasibility}

The gravitational energy gain due to the streaming of the gas from the virial
radius toward the centre of the halo potential well is a natural source of
energy for the \lya emission. The gravitational energy released per unit
infalling mass is of order $\Vv^2 = G\Mv/\Rv$, where the quantities are the
virial velocity, mass and radius respectively. The accretion rate $\dot{M}$ can
be estimated from the observed star-formation rate. For $V \sim 300 \kms$ and
$\dot{M} \sim 150 \sy$ we obtain a power $\dot{M}\, V^2 \sim 10^{43} \ergs$,
comparable to the luminosity of a typical LAB.  
 
In some more detail, an analysis of the {\MN} cosmological simulation
(described below in \se{sim}) reveals that both the cold gas accretion rate
$\dot M_{\rm c}$ and its inward velocity are roughly constant along the streams
\citep{nature}. If so, the gravitational power deposited at radius $r$ per unit
radial length in the cold gas is
\be
\dot E_{\rm grav} = f_{\rm c}\,\dot M_{\rm c}\, \left\vert \frac{\partial \phi}
{\partial r} \right\vert \, ,
\ee
where $\phi(r)$ is the gravitational potential at $r$. The factor $f_{\rm c}$
is the fraction of the energy that goes to heating the cold streams themselves
rather than the hot medium. The total power deposited between the virial radius
$\Rv$
and radius $r$, across a potential difference $\Delta \phi(\Rv,r)$, is
\be 
\dot E_{\rm grav} = f_{\rm c}\, \dot M_{\rm c} \, |\Delta \phi(\Rv,r)|\, .
\label{eq:heat1}
\ee

Given the total mass density profile $\rho(r)$ and the associated mass profile 
$M(r)$, the potential gain is
\be
|\Delta\phi| \equiv \fphi \Vv^2 , \quad
\fphi = -1 +\frac{V^2(r)}{\Vv^2}  
+\!\int_{{r}/{\Rv}}^1\!\!\! \frac{3\rho(r')}{\bar\rhov} r' \dd r' ,
\label{eq:phi}
\ee
where $V^2(r)=GM(r)/r$, $\bar\rhov$ is the mean density within the virial
radius, and $\fphi$ is typically a number of order unity.

For an NFW potential well with a concentration parameter $C$ \citep{nfw}, one
has 
\be 
\fphi(r)=\frac{C}{A_1(C)} \left[ \frac{\ln(1+x)}{x}-\frac{\ln(1+C)}{C} \right]
\, ,
\label{eq:fphi}
\ee
where $x=C r/\Rv$ and $A_1(x)=\ln (x+1) -x/(x+1)$. According to cosmological
$N$-body simulations \citep{bullock01}, the average concentration parameter as
a function of halo mass and redshift is $C \simeq 3\, M_{12}^{-0.13}
(1+z)_4^{-1}$, where $M_{12} \equiv \Mv/10^{12}\msun$ and $(1+z)_4\equiv (1+z)
/4$. For $\Mv=10^{12}\msun$, this is $C \simeq 3$ at $z=3$. Then $\fphi(r)
\simeq 4.7\,[\ln(1+x)/x-0.46]$ so $\fphi(x \rar 0) \simeq 2.5$.

A practical approximation for the average accretion rate into haloes of $\Mv
\sim\!  10^{12-13}\msun$ in the standard $\Lambda$CDM
cosmology\footnote{Adopting the parameters motivated by WMAP5: $\omm=0.28$,
$\oml=0.72$, $h=0.7$, $\sigma_8 = 0.8$.} is derived from the EPS approximation
and from the Millennium cosmological simulation
\citep[][Appendix A]{neistein06, neistein07, bdn07}. The same {\MN}
hydrodynamical simulation that is used in the current paper (\se{sim}) reveals
that 95\% of the gas accretion is through cold streams that penetrate
efficiently all the way into the vicinity of the central galaxy. For a baryonic
fraction of 0.165 in the incoming streams, the average cold accretion rate is
approximated by
\be
\dot M_{\rm c} \simeq 137 \sy\, M_{12}^{1.15}\, (1+z)_4^{2.25} \, .
\label{eq:acc}
\ee
This accretion rate is consistent with the typical SFR of $\sim 100\sy$
observed in massive galaxies of similar comoving number densities at redshifts
$2-3$ \citep{genzel06, foerster, elmegreen, genzel08, stark08}.

\begin{figure}
\begin{center}
\includegraphics[width=8.4cm]{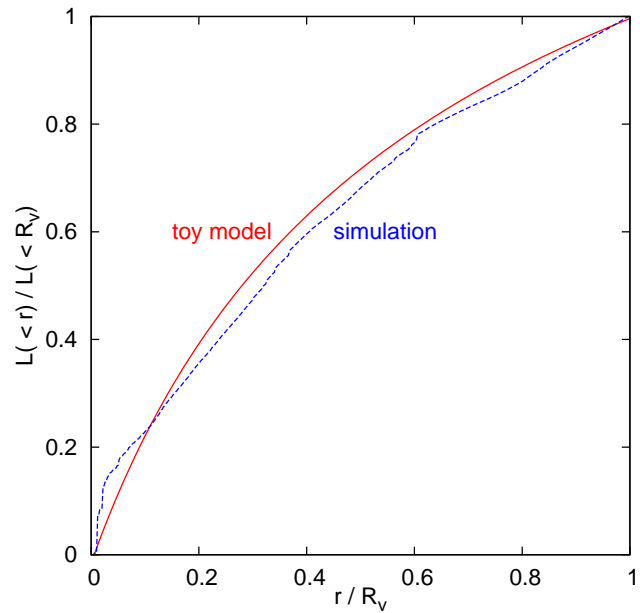}
\end{center}
\caption{Cumulative luminosity profile, showing the fraction of the luminosity
from within a sphere of radius $r$ compared to the virial sphere. The toy-model
prediction (solid red) is proportional to the fraction of gravitational energy
that is deposited inside radius $r$ based on \equ{eheatr}. The predicted
profile in the range $(0.2-0.5)\Rv$ can be approximated by the power law $L(<r)
\prop r^{0.7}$. About half the luminosity is expected to originate from the
cold streams in the halo outside the inner $0.27\Rv$ sphere. Shown in
comparison is the luminosity profile as derived from the three simulated {\CD}
galaxies stacked together (see section \ref{sec:grav}). The similarity between
the toy model predictions and the simulation results is remarkable.}
\label{fig:gravpot}
\end{figure}

The virial velocity is given by\footnote{With $\Rv \simeq 77\kpc\, M_{12}^{1/3}
\, (1+z)_4^{-1}$, for completeness.}
\be
\Vv \simeq 236\kms\, M_{12}^{1/3}\, (1+z)_4^{1/2} \, ,
\ee
so we finally obtain in \equ{heat1}
\be
\dot E_{\rm grav} \simeq 1.2\times 10^{43} \ergs\, f_{\rm c}\, 
M_{12}^{1.82}\, (1+z)_4^{3.25} \, .
\label{eq:eheat}
\ee
Assuming that a substantial fraction $\fa$ of this energy is emitted as
observable \lya radiation, we conclude that at $z \sim 3$, with $\fa \fc \sim
1$, a LAB of luminosity $L\sim 10^{43}\ergs$ is feasible from haloes of $\Mv
\sim 10^{12}\msun$. Luminosities as high as $\sim 10^{44}\ergs$ require haloes
of $\sim 3\times 10^{12}\msun$. If $\fa \fc$ is only $\sim 0.1$, then the
required haloes for the same luminosities are about three times more massive.
We note that the mean comoving number density for haloes more massive than 
$(1, 3, 10)\times 10^{12}\msun$ at $z=3$ is $(4.7, 0.68, 0.057)\times 10^{-4}
\Mpc^{-3}$ respectively. Given that LABs of a luminosity higher than
$10^{43}\ergs$ appear with a comoving number density of $\sim 5\times 10^{-5}
{\rm Mpc}^{-3}$ (\fig{st}), the simple gravitational heating model indicates
that they can emerge from haloes of $\sim 3\times 10^{12}\msun$, with $\fa \fc
\sim 0.1$. Thus, the comparison of the estimates from our toy model with the
total luminosities of the observed LABs indicates that gravitational heating is
a feasible source for the \lya emission. This kind of energy has to be released
from these galaxies.

One can combine equations (\ref{eq:fphi}) and (\ref{eq:eheat}) to evaluate the
gravitational energy deposited at different radii,
\begin{eqnarray}
\dot E_{\rm grav}(<r) & \simeq & 1.2\times 10^{43} \ergs\, f_{\rm c}\,
M_{12}^{1.82}\, (1+z)_4^{3.25} \nonumber \\ &\times&   \left[1.86 - {1.86\,
{\rm ln} (1 + C r / R_{\rm v}) \over C r / R_{\rm v}} \right] \, .
\label{eq:eheatr}
\end{eqnarray}
This gives a very crude estimate for the 3D luminosity profile $L(<r) = \fa \fc
\dot E_{\rm grav}(<r)$. This profile is shown in \fig{gravpot}, normalised to
the total luminosity inside the virial radius. One can read from this plot the
fraction of the luminosity that is expected in the different zones of the halo.
For example, about half the luminosity is expected from outside $r=0.27\Rv$,
implying that in haloes of $\Mv \sim 10^{12}\msun$ the LABs are expected to 
extend over more than $50\kpc$. The luminosity profile predicted by
gravitational heating is to be compared with our results from the simulations
(\se{grav}), and with observed LABs (\se{obs}).

\section{simulations}
\label{sec:sim}

We use here simulated galaxies from two different suites of simulations, both
employing Eulerian Adaptive Refinement Tree (AMR) hydrodynamics in a
cosmological setting. One suite consists of three simulations zooming in with a
maximum resolution of $35-70\pc$ on individual galaxies that reside in
dark-matter haloes of masses $\sim 5\times 10^{11}\msun$ at $z = 2.3$
\citep[][hereafter \CD]{cd}. The other suite is from the Horizon MareNostrum
simulation containing hundreds of massive galaxies in a cosmological box of
side $50\hmpc$ with a maximum resolution of $\simeq 1\kpc$  \citep[][hereafter
\MN]{ocvirk}. 

\begin{figure*}
\begin{center}
\includegraphics[width=7.64cm]{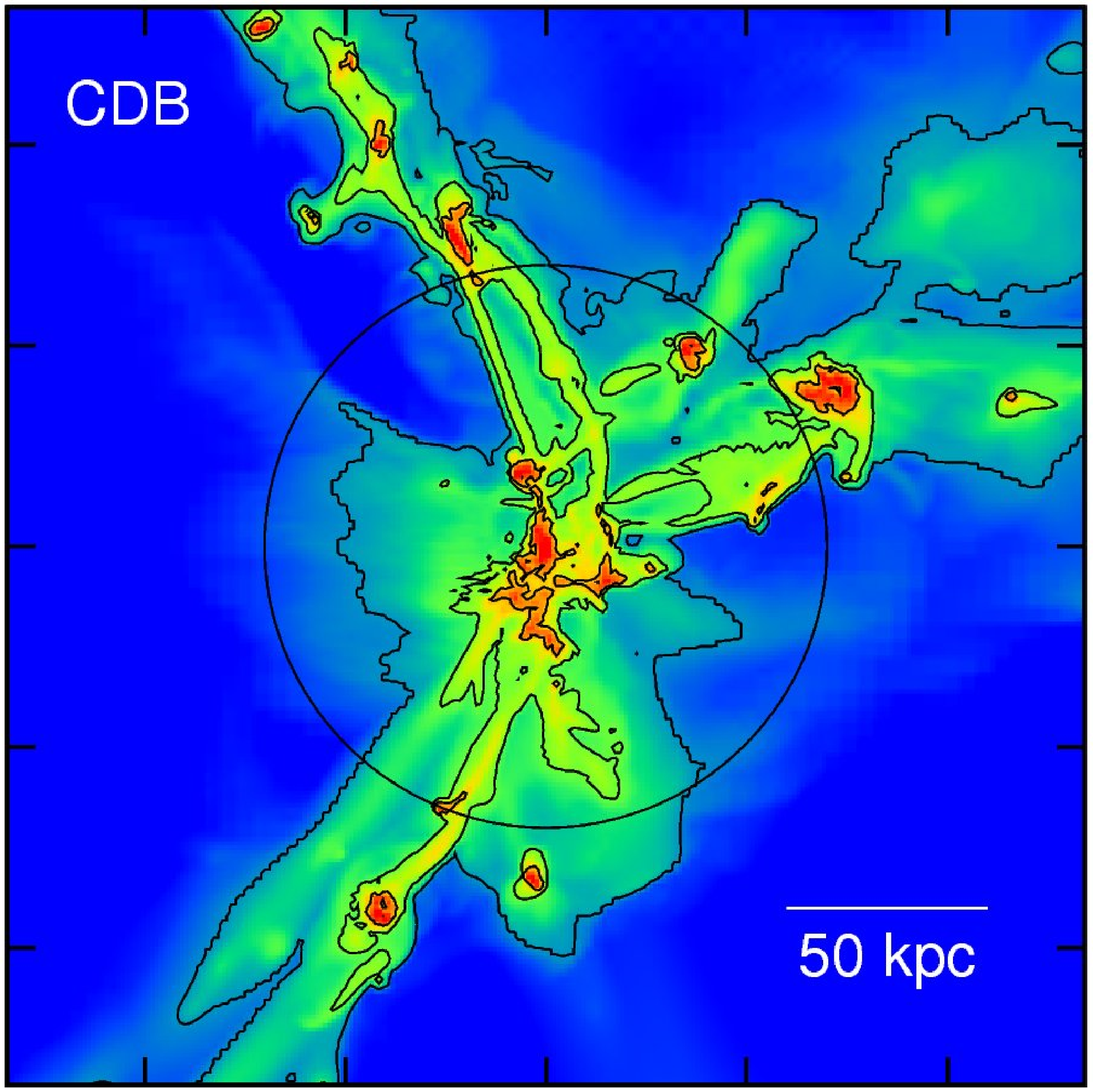}
\includegraphics[width=9.96cm]{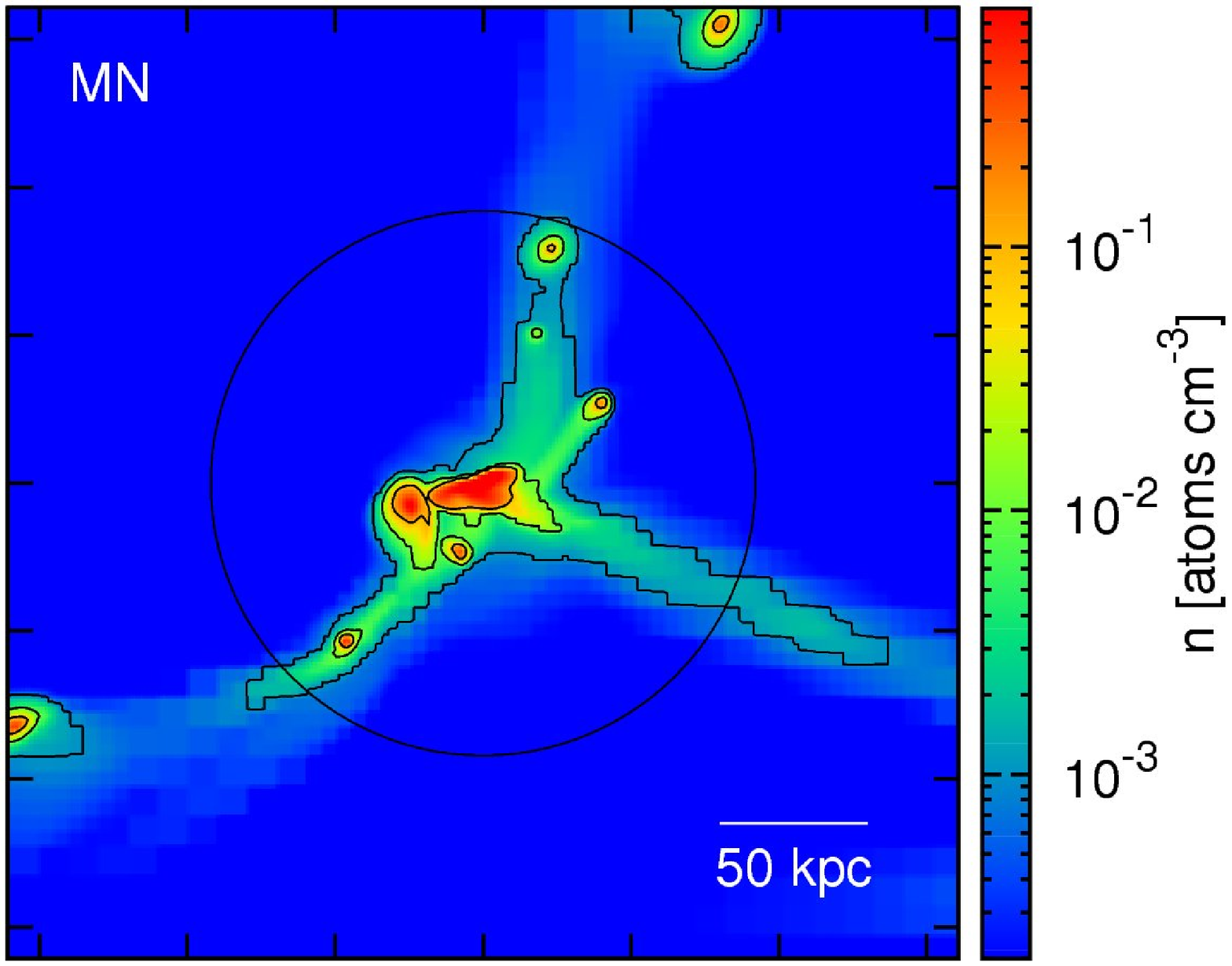}
\end{center}
\caption{Gas density in simulated galaxies from {\CD} (left) and {\MN} (right).
The colour refers to the maximum density along the line of sight. The contours
mark $n=0.1$, 0.01 and $0.001\cmmc$, respectively. The circle shows the virial
radius. Left: one of the three {\CD} galaxies (resolution $70\pc$) at $z=2.3$,
with $\Mv = 3.5 \times 10^{11}\msun$. Right: one of the {\MN} galaxies
(resolution $1\kpc$) at $z=2.5$, with $\Mv = 10^{12}\msun$. In both cases, the
inflow is dominated by three cold narrow streams that are partly clumpy. The
density in the streams is $n=0.003-0.1 \cmmc$, with the clump cores reaching $n
\sim 1\cmmc$.}
\label{fig:denmap}
\end{figure*}

\Fig{denmap} shows sample gas density maps of galaxies from the two suites of
simulations. They demonstrate the dominance of typically three, narrow cold
streams, which come from well outside the virial radius along the dark-matter
filaments of the cosmic web, and penetrate into the discs at the halo centres.
The streams are partly clumpy and partly smooth, even in the simulation of
higher resolution. The typical densities in the streams are in the range $n=
0.01-0.1\cmmc$, and they reach $n=0.1-1\cmmc$ at the clump centres and in the
central disk.

\subsection{High-resolution ART CDB simulations}

The {\CD} simulations have been run with the AMR tree code \textsc{Art}
\citep{kkk,andrey} with a spatial resolution better than $70 \pc$ in physical
units. The code incorporates the relevant physical processes for galaxy
formation, including gas cooling and photoionisation heating, star formation,
metal enrichment and stellar feedback \citep{cak}. Cooling rates were computed
for the given gas density, temperature, metallicity, and UV background based on
\textsc{Cloudy} \citep{ferland}, assuming cooling at the centre of a cloud of
thickness 1 kpc \citep{ceverino, rk}. Metals are included, assuming an
abundance of 0.02 relative to the total mass for a solar composition. The code
implements a ``constant" feedback model, in which the combined energy from
stellar winds and supernova explosions is released as a constant heating rate
over 40 Myr, the typical age of the lightest star that explodes as a type-II
supernova. Photo-heating is also taken into account self-consistently with
radiative cooling. A uniform UV background based on the \citet{haardt} model is
assumed, local sources are ignored. In order to mimic the self-shielding of
dense, galactic neutral hydrogen from the cosmological UV background, the
simulation assumes a substantially suppressed UV background ($5.9 \times
10^{26} \ {\rm erg} \ {\rm s}^{-1} \ {\rm cm}^{-2} \ {\rm Hz}^{-1}$, the value
of the pre-reionisation UV background at $z=8$) for the gas at total gas
densities above $n=0.1$ cm$^{-3}$.

The unique feature of this code for the purpose of simulating the detailed
structure of the streams and the gravitational instability in the disk is to
allow the gas to cool down well below $10^4$K, thus reaching high densities in
pressure equilibrium with the hotter and more dilute medium. A non-thermal
pressure floor has been implemented to ensure that the the Jeans length is
resolved by at least seven resolution elements and thus prevent artificial
fragmentation on the smallest grid scale \citep{truelove,rk,cd}. It is
effective in the dense $(n > 10$ cm$^{-3})$ and cold $(T<10^4 K)$ regions inside
galactic disks, while most of the \lya emitting gas, which is at temperatures
$(1-5)\times 10^4$K, is not affected by this pressure floor.

The equation of state remains unchanged at all densities. Stars form in cells
where the gas temperature is below $10^4$K and its density is above the
threshold $n = 1\cmmc$ according to a stochastic model that is roughly
consistent with the \citet{kennicutt} law. The ISM is enriched by metals from
supernovae type II and type Ia. Metals are assumed to be released from each
stellar particle by SNII at a constant rate for 40 Myr since its birth,
assuming a \citet{miller} IMF and matching the results of \citet{woosley}. The
metal ejection by SNIa assumes an exponentially declining SNIa rate from a
maximum at 1 Gyr. The code treats the advection of metals self-consistently
\citep{ceverino}.

The dark matter particle mass is $5.5 \times 10^5 \msun$, the minimum star
particle mass is $10^4 \msun$, the smallest cell size is 70 pc (physical
units), and the force softening length is 105 pc.

The initial conditions for the {\CD} simulations were created using a
low-resolution cosmological $N$-body simulation in a comoving box of side $20
\hmpc$, for which the cosmological parameters were motivated by the WMAP5
following values \citep{WMAP5}: $\Omega_{\rm m} = 0.27$, $\Omega_\Lambda =
0.73$, $\Omega_{\rm b} = 0.045$, $h = 0.7$ and $\sigma_8 = 0.82$. At $z=1$,
three haloes of $\Mv \simeq 10^{12}\msun$ each have been selected, avoiding
haloes that were subject to a major merger near that time. The three halo
masses at $z=2.3$ are $3.5, 4, 6\times 10^{11}\msun$, and they end up as
$(3-4)\times 10^{12}$ M$_\odot$ haloes today. For each halo, a concentric
sphere of radius twice the virial radius was identified for re-simulation with
high resolution. Gas was added to the box following the dark matter
distribution with a fraction $f_{\rm b} = 0.15$. The whole box was then
re-simulated, with refined resolution only in the selected volume about the
respective galaxy.

\subsection{RAMSES Mare Nostrum simulations}

The {\MN} simulation uses the AMR code \textsc{Ramses} \citep{teyssier}. The
spatial resolution is $\sim 1$kpc in physical units. UV heating is included
assuming the \citet{haardt} background model, as in the {\CD} simulation. The
code incorporates a simple model of supernovae feedback and metal enrichment
using the implementation described in \citet{dubois}. The cooling rates are
calculated assuming ionisation equilibrium for H and He, including both
collisional- and photo-ionisation \citep{katz}. Metal cooling is also included
using tabulated \textsc{Cloudy} rates, and is assumed proportional to the
metallicity, relative to the \citet{grevesse} solar abundances. Unlike in the
{\CD} simulation, no cooling below $T<10^4$K is computed, and no self-shielding
of the UV flux is assumed. Because we assume that the \lya emissions are
produced in predominantly shielded gas that is in CIE, our {\MN} computations
should be regarded as less accurate than the {\CD} simulation.

For high-density regions, the MN code considers a polytropic equation of state
with $\gamma_0 = 5/3$ to model the complex, multi-phase and turbulent structure
of the inter-stellar medium (ISM) \citep{yepes, sh} in a simplified form
\citep[see][]{joop, dubois}. The ISM is defined as gas with hydrogen density
greater than $n_{\rm H} = 0.1\cmmc$, one order of magnitude lower than in the
{\CD} simulation. Star formation has been included, for ISM gas only, by
spawning star particles at a rate consistent with the \citet{kennicutt} law
derived from local observations of star forming galaxies. 

The {\MN} simulation implemented a pressure floor in order to prevent
artificial fragmentation, by keeping the Jeans lengthscale, $\lambda_{\rm J}
\prop T n^{-2/3}$, larger than 4 grid-cell sides everywhere. In any case where
$n>0.1\cmmc$, a density dependent temperature floor was imposed. It mimics the
average thermal and turbulent pressure of the multiphase ISM, in the spirit of
\citet{sh} or \citet{dvs}. In our case, we allow the gas to heat up above this
temperature floor and cool back. The temperature floor follows a polytropic
equation of state with $T_{\rm floor}=T_0 (n/n_0)^{\gamma_0-1}$, where
$T_0=10^4$ K and $n_0=$ 0.1 atoms cm$^{-3}$. The resulting pressure floor is
given by $P_{\rm floor} = n_{\rm H} k_{\rm B} T_{\rm floor}$.

We crudely correct for this artificial temperature boost in post-processing by
subtracting $T_{\rm floor}$ from the temperature read from the grid cells where
$n>0.1\cmmc$. If the corrected temperature is below $10^4$K, we set it to
$10^4$K. In practise, almost all cells where $n>0.1 \cmmc$ are set to $T =
10^4$K. As will become clear in \se{lal}, this means neglecting any \lya
emission from these high-density cells. We will evaluate the possible error
made by this procedure in the {\MN} galaxies using the more accurate
high-resolution {\CD} galaxies, where no temperature floor has been applied.

For each stellar population, 10\% of the mass is assumed to turn into
supernovae type II after 10 Myr, where the energy and metals are released in an
impulse. For each supernova, 10\% of the ejected mass is assumed to be pure
metals, with the remaining 90\% keeping the metallicity of the star at birth.
SNIa feedback has not been considered.

The dark matter particle mass is $1.16 \times 10^7 \msun$, the star particle
mass s $2.05 \times 10^6 \msun$, the smallest cell size is $1.09$ kpc physical,
and the force softening length is  1.65 kpc.

The initial conditions of the \MN simulation were constructed assuming a 
$\Lambda$CDM universe with $\Omega_{\rm M} = 0.3$, $\Omega_\Lambda = 0.7$, 
$\Omega_{\rm b} = 0.045$, $h=0.7$ and $\sigma_8 = 0.9$ in a periodic box 
of $50\hmpc$. The adaptive-resolution rules in this simulation were the same
everywhere, with no zoom-in resimulation of individual galaxies.

\section{Computing the \lya Luminosity}
\label{sec:lal}

\begin{figure}
\begin{center}
\includegraphics[width=8.4cm]{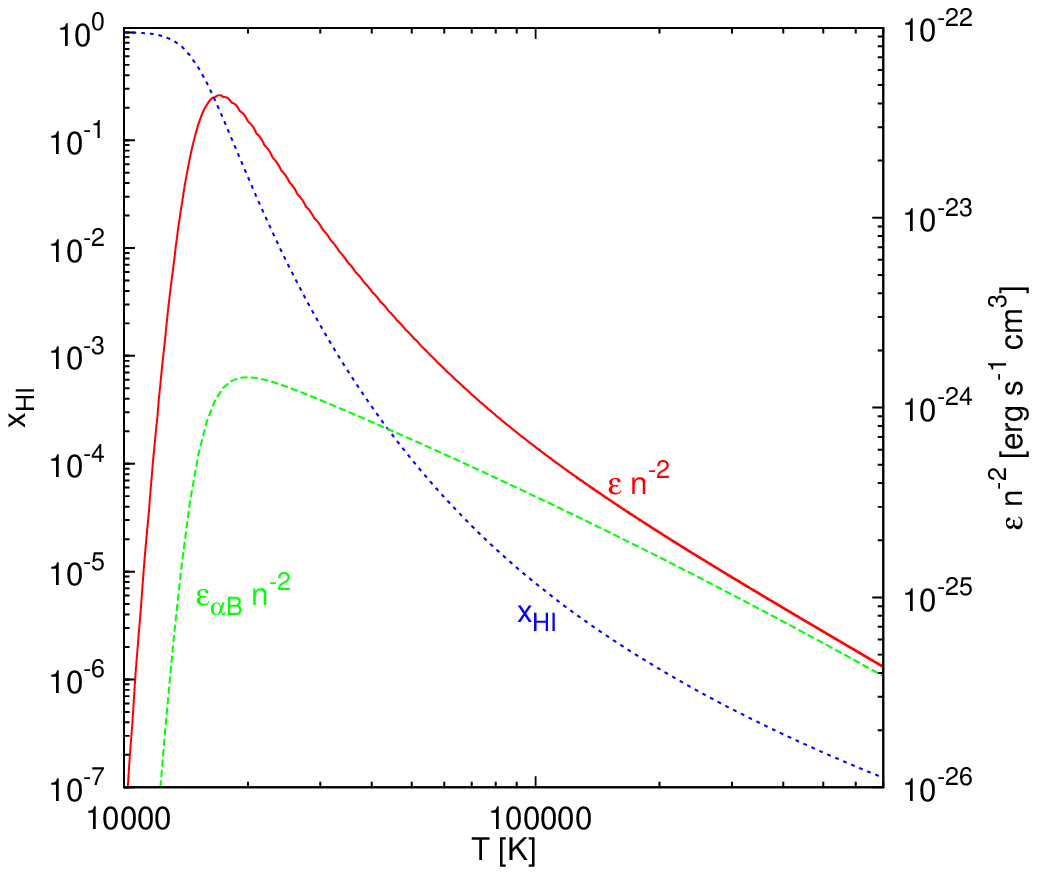}
\end{center}
\caption{Temperature dependence of the ionisation state and the \lya
emissivity. The neutral-hydrogen fraction $x_{\rm HI}$ (short-dashed, blue
curve) is derived from \equ{xHI} assuming CIE. The temperature dependence of the
emissivity $\epsilon$ (solid, red curve) is computed from \equs{eps},
\equm{nHI} and \equm{nHII} for $n=1\cmmc$. We see that the gas is neutral for
$T \leq 10^4$K and it becomes highly ionised at $T>2\times 10^4$K. The maximum
emissivity at a given density is obtained at $T \simeq 1.8 \times 10^4$K, and
the emissivity is high enough to substantially contribute to an overall
luminosity in the range $T = (1 - 5) \times 10^4$K. The contribution of
recombination to the emissivity, $\epsilon_{\alpha {\rm B}}$ (long-dashed, green
curve), is negligible in the relevant temperature range.}
\label{fig:xHIelyamax}
\end{figure}

Given the gas temperature $T$ and mass density $\rho$ in every cubic grid cell
of the hydrodynamic simulation, we compute the local \lya emissivity produced
by electron impact excitation of neutral hydrogen,
\be
\epsilon =  n_{\rm e}\, n_{\HI}\, q_{1s\rar 2p}(T)\, h \nu_{\La}\  
{\rm erg\,cm^{-3}\,s^{-1}} \, .
\label{eq:eps}
\ee
In this expression, $n_{\rm e}$ and $n_{\HI}$ are the local electron and atomic
hydrogen particle densities (in $\cmmc$) and $h\nu_{\La} = 10.2\, {\rm eV} =
1.63\times 10^{-11}\,{\rm erg}$ is the \lya photon energy. The temperature
dependent quantity $q_{1s\rar 2p}$ is the collisional excitation rate
coefficient \citep{callaway87},
\bea
q_{1s\rar 2p} &=& \frac{2.41\times 10^{-6}}{T^{0.5}}\,
\left(\frac{T}{10^4}\right)^{0.22} \nonumber \\ 
& \times &\exp \left(-\frac{h\nu_{La}}{kT}\right) \ {\rm cm^3\, s^{-1}} \, ,
\label{eq:q}
\eea
where $T$ is in degrees K. Radiative recombinations of electrons with protons
contribute negligibly to the \lya emissivity, as can be seen from the green,
long-dashed curve in Figure \ref{fig:xHIelyamax}.

We assume a primordial helium mass fraction $Y=0.24$, corresponding to a helium
particle abundance of $1/12$ relative to hydrogen. This gives
\be
\rho = (4/3)\, m_{\H}\, n_{\H} \, ,
\label{eq:rho}
\ee
where $m_{\H}$ is the hydrogen mass and $n_{\H}$ is the total density of
hydrogen nuclei (neutral plus ionised). The electron and atomic hydrogen
densities in \equ{eps} are given by
\be
n_{\HI}/n_{\H} = x_{\HI}
\label{eq:nHI}
\ee
and
\be
n_{\rm e}/n_{\H} = x_{\HII} + (1/12)\,(x_{\HeII} + 2 x_{\HeIII}) \, ,
\label{eq:nHII}
\ee
where we adopt the temperature dependent ionisation fractions, $x_{\HI}$,
$x_{\HII}$, $x_{\HeII}$ and $x_{\HeIII}$ as computed for collisional ionisation
equilibrium (CIE) assuming case-B hydrogen recombination \citep[O.~Gnat,
private communication; see also][]{gnat}. A large fraction ($\sim 95\%$) of the
\lya emission is produced in gas where the helium is fully neutral. In this
limit,
\be
x_{\HI} = \alpha_{\rm B} / (\alpha_{\rm B} + q_{\rm i}) \, ,
\label{eq:xHI}
\ee
where $\alpha_{\rm B}$ and $q_{\rm i}$ are the recombination and collisional
ionisation rate coefficients \citep{gnat}.

We note for completeness that approximate fits can be provided by
\be
\alpha_{\rm B}(T) = 4.9 \times 10^{-6}\, T^{-1.5}
\left(1 + \frac{115}{T^{0.41}}\right)^{-2.24}\cmpc\ {\rm s}^{-1} 
\ee
and
\begin{eqnarray}
q_{\rm i} & = & 21.11\, {\rm cm^3\, s^{-1}\, K^{3/2}}\, T^{-3/2}\, {\rm exp}
\left(-{T_{\rm HI} \over T} \right) \nonumber \\
& \times & {(2T_{\rm HI}/T)^{-1.089} \over \left[1 + \left(5.65\, T_{\rm HI} /
T\right)^{0.874}\right]^{1.101}}\, ,
\end{eqnarray}
with $T_{\rm HI} = 1.58\times 10^5$K being the ionisation threshold according
to \citet{hui}. We do not use these fits in our calculation here.

We assume that the cold streams are practically {\it self-shielded\,} against
the photoionising UV background, for the following reason. The typical HI
column densities along the shortest dimensions of the streams are $\sim 10^{20}
\cmms$ and the total hydrogen particle densities are in the range $0.01-0.1
\cmmc$, or typically $\sim 0.03 \cmmc$ (see \fig{nhistod}). At redshift $z=3$,
the mean Lyman continuum photon intensity is $4\pi J^* \approx 2.2\times 10^5
{\rm photons\ s^{-1}\, cm^{-2}}$ \citep{haardt} and the unattenuated hydrogen
photoionisation rate is $\Gamma = 5.6\times 10^{-13}\,{\rm s^{-1}}$. Thus, for
gas densities $n \lsim 2\Gamma/\alpha_B \simeq 4.3\,{\rm cm^{-3}}$, the gas
will be more than 50\% ionised by the unattenuated field. However, absorption
of the radiation field in the stream gas will produce a photoionised column
\be
N^{\rm photo}_{\II} = \frac{2\pi J^*}{n_{\H}\, \alpha_B} 
= \frac{4.2\times 10^{17}}{n_{\H}}\, \cmms \, .
\label{eq:shielding}
\ee
For the typical stream volume densities this is small compared to the neutral
columns of $\sim 10^{20} {\rm cm}^{-2}$ that we find, so the streams can be
practically assumed to be self-shielded against the background radiation.
Parts of the streams may not be self-shielded against local sources of
radiation within the streams, e. g. from star-forming satellites. In those case
we might overpredict the HI fraction and thus the \lya luminosity.

The CIE neutral hydrogen fraction as a function of temperature and the
temperature dependence of the \lya emissivity for a given total density are
shown in \fig{xHIelyamax}. One can see that the gas is neutral at $T=10^4$K and
below, and it becomes highly ionised at $T>2\times 10^4$K. The maximum
emissivity at a given density is obtained at $T \simeq 1.8 \times 10^4$K, and
the emissivity is high enough to substantially contribute to the overall
luminosity in the range $T = (1 - 5) \times 10^4$K. Hydrogen in this
temperature range is thus the expected source of \lya emission.

\begin{figure}
\begin{center}
\includegraphics[width=8.4cm]{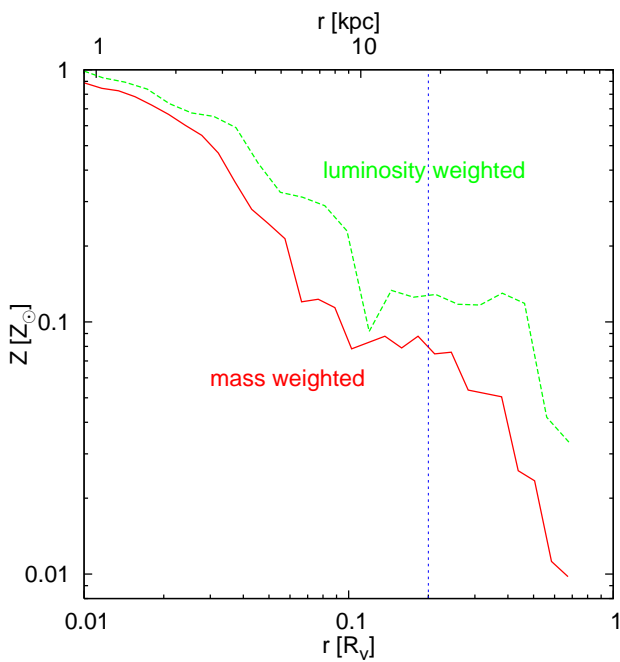}
\end{center}
\caption{Average metallicity profile in the simulated {\CD} galaxies, once
weighted by the density of the \lya emitting cold gas in the temperature range
$(1-5) \times 10^4$K and once weighted by the luminosity. The vertical line
indicates 0.2 $R_{\rm v}$ which we refer to as the outer border of the galaxy.}
\label{fig:Z}
\end{figure}

The total \lya luminosity from a given volume is assumed to be a constant
fraction of the sum over all the cells that are contained in that volume, 
\begin{equation}
L_{\La} = \fa \sum_i \epsilon_i\, V_i \, ,
\label{eq:lum2}
\end{equation}
where $V_i$ is the cell volume. The complex radiation transfer process along
the cosmological line of sight is summed up in the transmission factor $\fa$,
the fraction of the \lya photons that make it to the observer. In \se{lumfun},
we will determine the actual value of $\fa$ by matching the predictions from
the simulations with a preliminary determination of the observed \lya
luminosity function. We expect $\fa$ to be in the range 0.5-1. This is because
the main source of opacity is likely to be intervening and lower redshift
intergalactic HI, which is expected to absorb part of the blue side of the line
profile. An estimate for a typical line of sight to $z\sim 3$ is $\fa \sim
0.85$ \citep[e.g.][]{faucherb}. The transmission factor along the line of sight
to a typical LAB could be slightly smaller because the LABs tend to reside in
overdense environments. A smooth component of HI in the emitting halo may
absorb some of the red wing as well. On the other hand, as estimated next,
dust opacity is expected to reduce $\fa$ by a small factor only except in the
galaxy itself and in its immediate vicinity.

Given the gas metallicity in every cell of the simulation, we can estimate the
dust absorption as follows. For a medium in which the dust abundance scales
linearly with the metallicity $Z$ (with $Z \approx 1$ corresponding to Galactic
dust), the continuum opacity due to dust absorption at the \lya wavelength is
$\tau_{\rm dust} \simeq  0.1\, Z\, N_{20}$ where $N_{20}$ is the
neutral-hydrogen column density in units of $10^{20}\cmms$, \citep{draine03}.
The line optical depth is $\tau_{\La} \simeq 10^7\, N_{20}\, T_4^{-1/2}$
\cite{neufeld} so that $\tau_{\rm dust} \simeq 10^{-8}\,Z\, T_4^{1/2}\,
\tau_{\La}$. For \lya scattering through a plane-parallel medium the total line
optical depth traversed is $\sim (9\, a\, \tau_0 / \sqrt{\pi})^{1/3}\tau_0$,
where $\tau_0$ is the \lya optical depth from the mid-plane to the surface, and
$a= 4.7\times 10^{-4}\,T_4^{-1/2}$ is the damping constant \citep{adams75}.
Thus, as the photon scatters through the medium the effective dust opacity is
\be
\taudust \simeq 2.9\,Z\, N_{20}^{4/3}\, T_4^{-1/3} \, .
\label{eq:taudust}
\ee
For $Z = 0.1$, assuming $T_4 \sim 1$, we obtain $\tau_{\rm dust} = 1$ for
$N_{20} \simeq 3$. We find in the CDB galaxies that the \lya luminosity-weighted
fraction of volume elements that have a column density $N_{20} \ge 3$ between
them and the observer, averaged over different directions, is less than a third.
This provides an estimate of no more than a third reduction outside $r \sim 0.1
\Rv$ (where $Z \le 0.1$).

\Fig{Z} shows the stacked 3D metallicity profile of the \lya emitting gas in 
the three simulated \CD galaxies. It is computed for gas in the temperature 
range $(1-5)\times 10^4$K and is once weighted by gas density and once by
luminosity. The metallicity profiles were computed in spherical shells. The
profile shows that the metallicity falls below 0.1 solar outside the disc
radius of $\sim 8\kpc$, and it drops to much smaller values outside the inner
$\sim 30\kpc$, where the streams are basically made of compressed intergalactic
gas. Therefore dust opacity is estimated to reduce the \lya luminosity from the
streams by a small factor only. However, this assumption is likely to fail in
the inner galactic disk, where the metallicity is above 0.1 solar despite the
high redshifts. The effective transmission parameter $\fa$ in the galaxy
vicinity is lower than in the streams. By ignoring the dust, we overestimate
the \lya luminosity from the disk.

\section{The Lyman-Alpha Emitting Gas}
\label{sec:lyagas}

\begin{figure}
\begin{center}
\includegraphics[width=8.4cm]{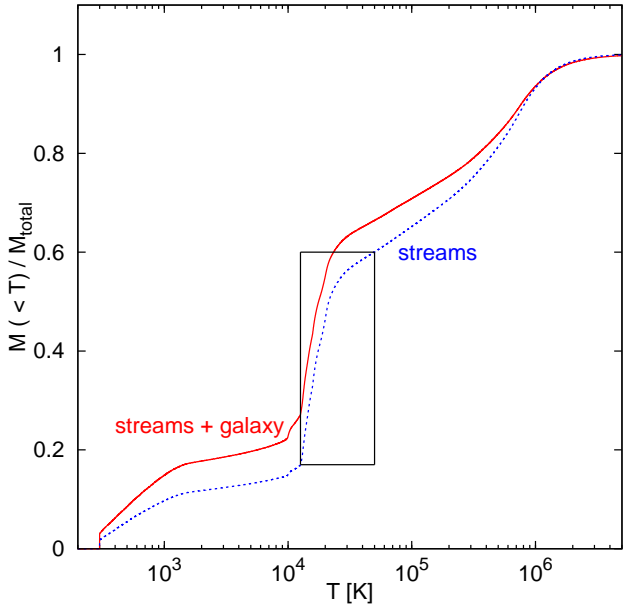}
\end{center}
\caption{Cumulative, mass-weighted temperature distribution in the stacked 3
{\CD} haloes at $z = 2.3$. The curves refer to the streams at $r=(0.2-1)\Rv$
(dashed blue) and to the whole halo including the inner galaxy (solid red). We
see that the temperature distribution throughout the halo is bimodal, with cold
and hot phases at $10^4<T<3\times 10^4$K and $3\times 10^5 < T < 2 \times
10^6$K, respectively. Outside the inner $0.2\Rv$, the hot medium fills the halo
while the cold gas is confined to inflowing narrow streams. The cold tail at
$T<10^4$K comes from dense clumps within the streams. The box marks the
temperature range of the \lya-emitting gas.}
\label{fig:thistod}
\end{figure}

\begin{figure}
\begin{center}
\includegraphics[width=8.4cm]{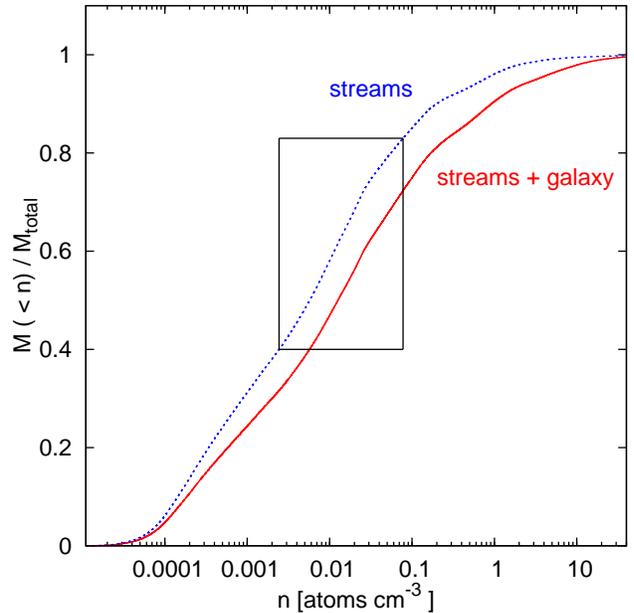}
\end{center}
\caption{Cumulative, mass-weighted gas density distribution in the stacked 3
{\CD} haloes at $z = 2.3$. The curves refer to the streams at $r=(0.2-1)\Rv$
(dashed blue) and to the whole halo including the inner galaxy (solid red). The
density in the cold streams is in the range $n = 0.001 - 0.1 \cmmc$, and in the
denser colder clumps it becomes $n=0.1-1\cmmc$. The box marks the density range
of the \lya-emitting gas, based on the box marked in \fig{thistod} and assuming
that $n$ is a monotonically decreasing function of $T$.}
\label{fig:nhistod}
\end{figure}

\Fig{thistod} shows the cumulative mass-weighted temperature distribution in
the {\CD} galaxies at $z=2.3$. The distribution is specified alternatively for
the halo gas in the radius range $(0.2-1.0)\Rv$ and for the whole gas,
including the inner part that involves the disc and its neighbourhood. We see
that the temperature distribution in the halo is bimodal, with a hot virial
phase at $10^5 < T < 2 \times 10^6$K containing $\sim 35\%$ of the gas and a
cold phase (marked by a box) at $10^4<T<3\times 10^4$K containing $\sim 50\%$
of the gas. A very cold tail at $T<10^4$K contains $\sim 15\%$ of the gas. For
the somewhat more massive {\MN} galaxies, which we do not show here, the
situation is qualitatively similar. The temperature distribution is again
bimodal, with a hot phase at $6 \times 10^4 < T < 5 \times 10^6$K containing
$\sim 45\%$ of the gas and a cold phase at $2 \times 10^4 < T < 4 \times 10^4$K
containing $\sim 35\%$ of the gas. An intermediate component in the range $4
\times 10^4 < T < 6 \times 10^4$K contains $\sim 20\%$ of the gas. In the two
kinds of simulations, while the hot phase is spread throughout the halo, the
cold phase is in the narrow streams flowing through the hot medium and the very
cold tail is concentrated in the dense clumps within the streams and in the
disc. Based on the temperature dependence of the emissivity in
\fig{xHIelyamax}, we conclude that most of the \lya luminosity is expected to
come from the cold streams, with about half the total gas mass participating in
efficient \lya emission.

\begin{figure*}
\begin{center}
\includegraphics[width=8.93cm]{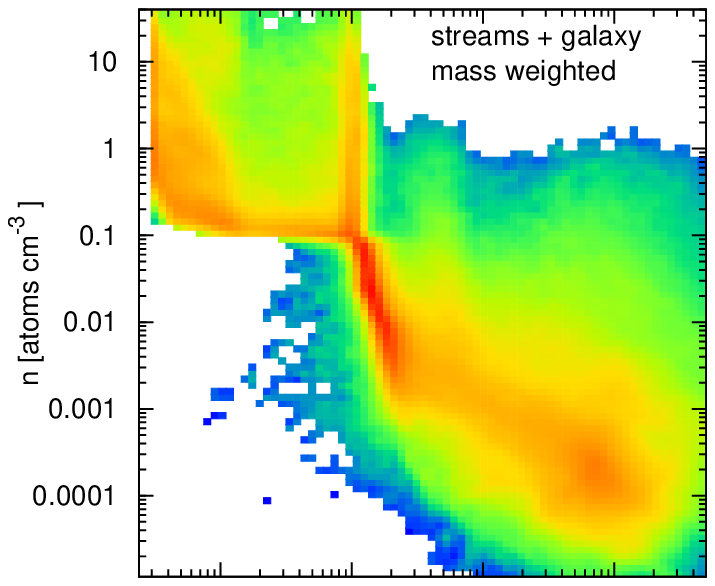}
\includegraphics[width=8.67cm]{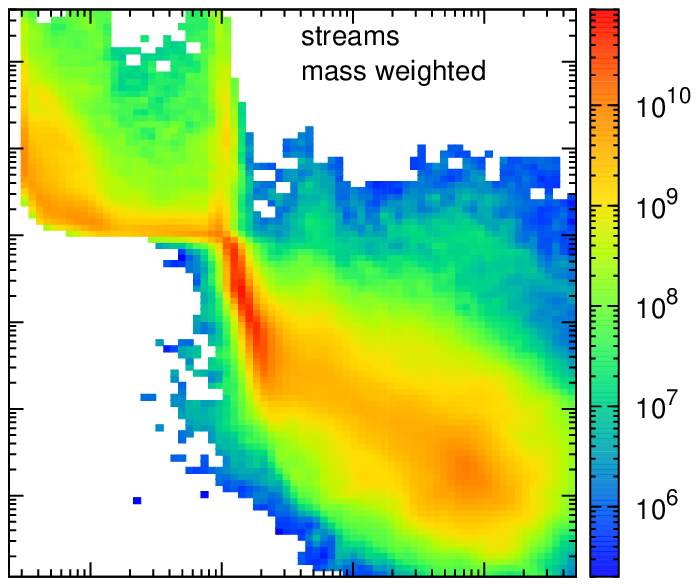}
\includegraphics[width=8.93cm]{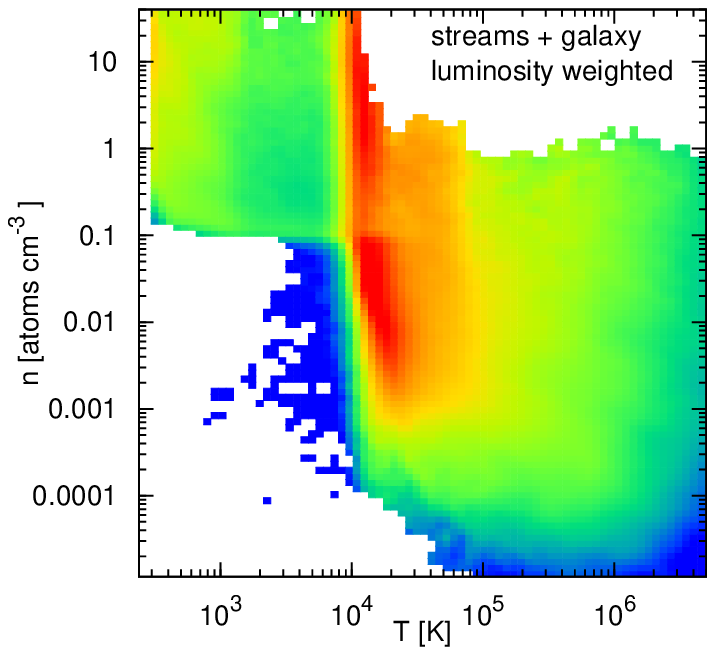}
\includegraphics[width=8.67cm]{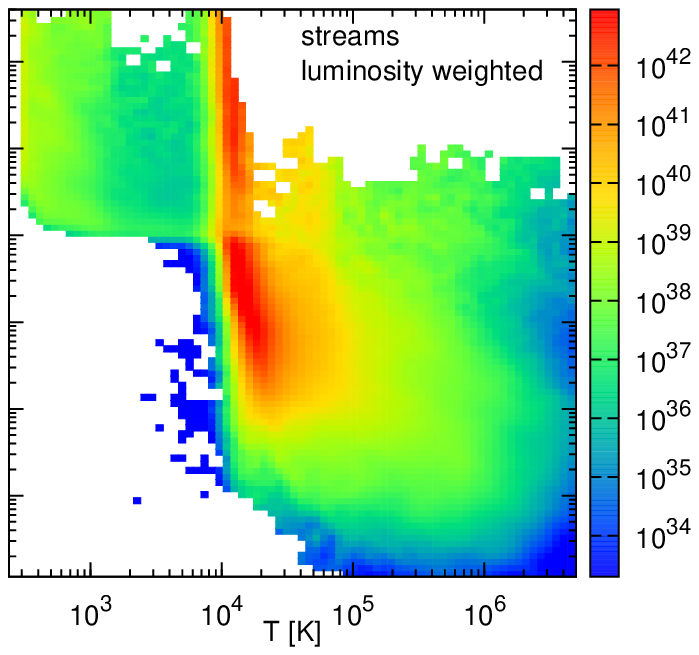}
\end{center}
\caption{Distribution of mass (top) and luminosity (bottom) in the $n-T$ plane
for the stacked 3 {\CD} haloes at $z = 2.3$. The panels refer to the streams at
$r=(0.2-1)\Rv$ (right) and to the whole halo including the inner galaxy (left).
The units are mass M$_\odot$ (top) and erg s$^{-1}$ (bottom) per unit area in
the (log$_{10}$ $n$, log$_{10}$ $T$) plane. The luminosity peaks at
$T=(1-2)\times 10^4$K over a broad range of densities from $n=0.001 \cmmc$ and
up. Most of the emission is from the streams outside the central galaxy.}
\label{fig:nthistod}
\end{figure*}

\Fig{nhistod} shows the cumulative mass-weighted gas density distribution in
the {\CD} galaxies at $z=2.33$ in the same radius zones. If the multi-phase
medium is in pressure equilibrium, the rank order of the cells by density and
by temperature are exactly opposite of one another, so a comparison of
\fig{thistod} and \fig{nhistod} reveals that the \lya-emitting cold streams
have densities in the range from below $0.01$ to above $0.1 \cmmc$ (marked by a
box). The situation for the {\MN} galaxies is comparable.

\Fig{nthistod} shows the two-dimensional gas distribution in the
temperature-density plane. It complements Figures \ref{fig:thistod} and
\ref{fig:nhistod} with information on the different gas phases present in the
simulations. We see that much of the gas is not in pressure equilibrium,
$n T = {\rm const}$. The top and bottom panels display the distribution of mass
and \lya luminosity, respectively. We see that the luminosity peaks at
$T=(1-2)\times 10^4$K over a broad range of densities from $n=0.001 \cmmc$ and
up. We interpret the $T>10^4$K gas with densities above $n=0.1 \cmmc$ outside
the central galaxy to be in clumps along the streams. It seems to be associated
with a few percent of the mass that contribute about 25\% of the luminosity. A
similar fraction of the luminosity comes form gas in the temperature range
$T=(2-7)\times 10^4$K.

In order to evaluate the possible error introduced in the {\MN} galaxies by
practically ignoring the emissivity from cells with densities $n>0.1\cmmc$,
we computed the fraction of the luminosity coming from such cells in the {\CD}
galaxies. We find that this fraction is typically only $\sim 15\%$ in the halo
at $r>0.2\Rv$, and it could be as high as 40\% in the central disc. By adding
15\% to the \MN emissivities everywhere we obtain a good match between the
total luminosities at a given halo mass in the \CD and \MN simulations (see
\fig{lumiz246}).

\section{Simulated Lyman-alpha Blobs}
\label{sec:images}

\begin{figure*}
\begin{center}
\includegraphics[width=7.07cm]{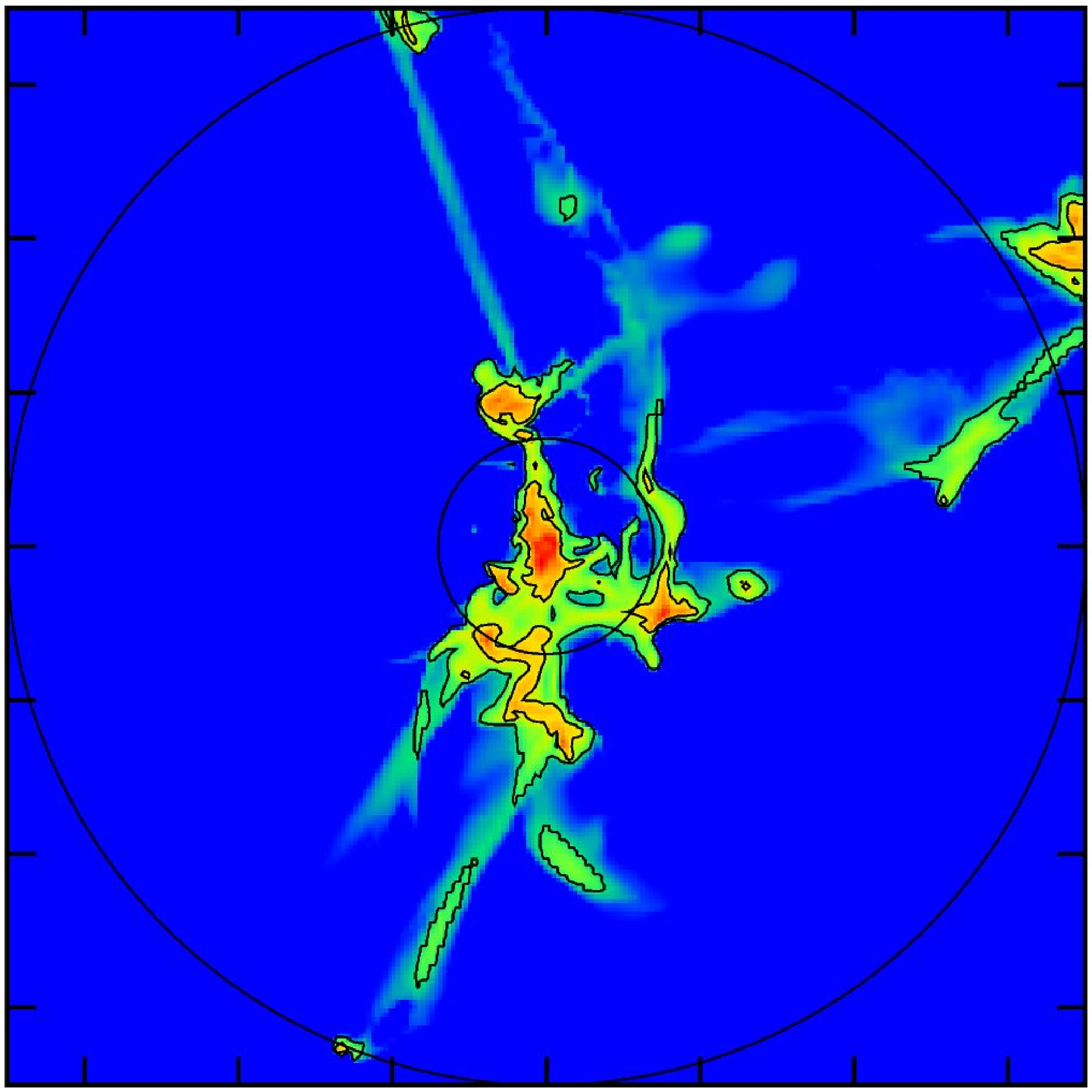}
\includegraphics[width=9.27cm]{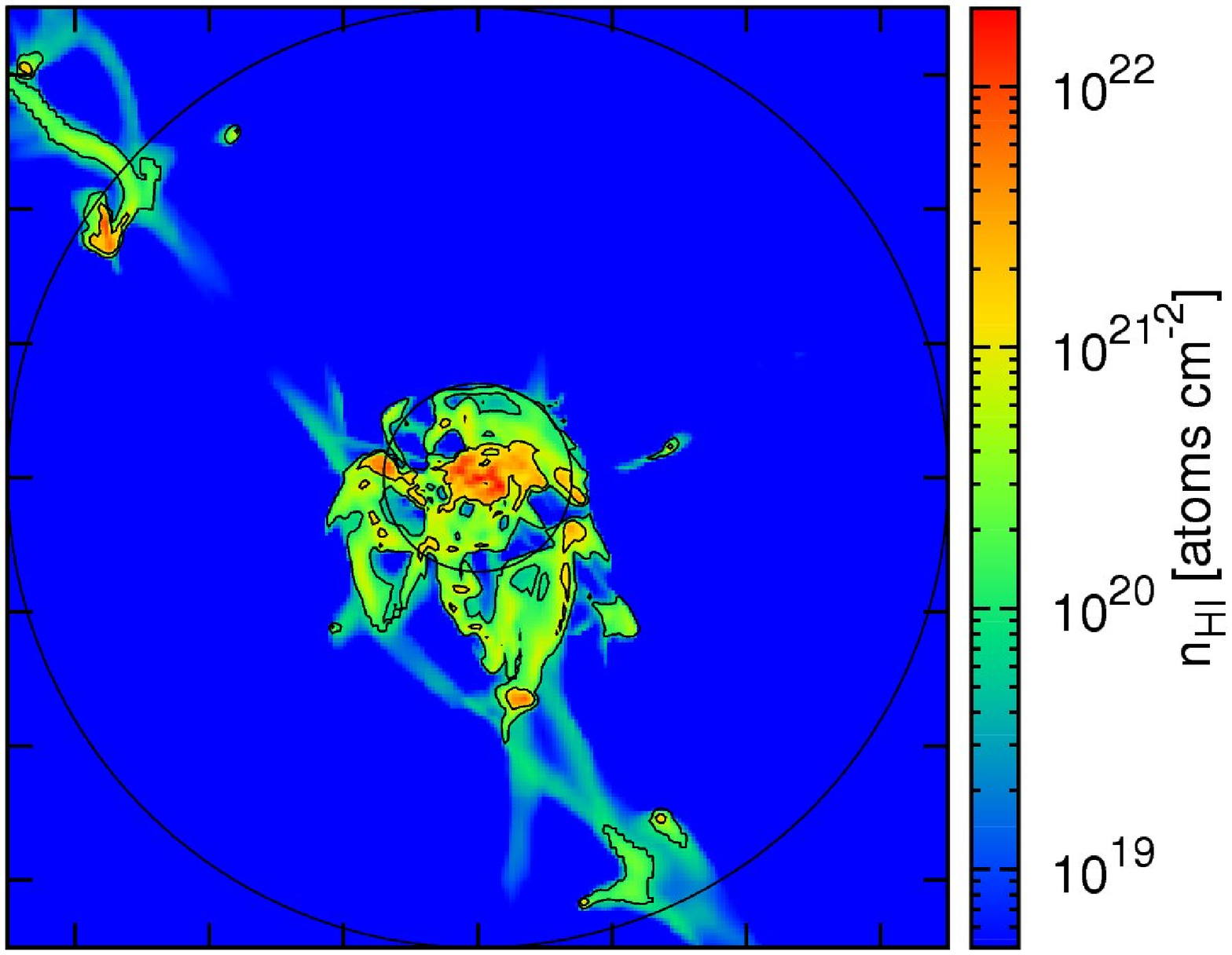}\\
\includegraphics[width=7.07cm]{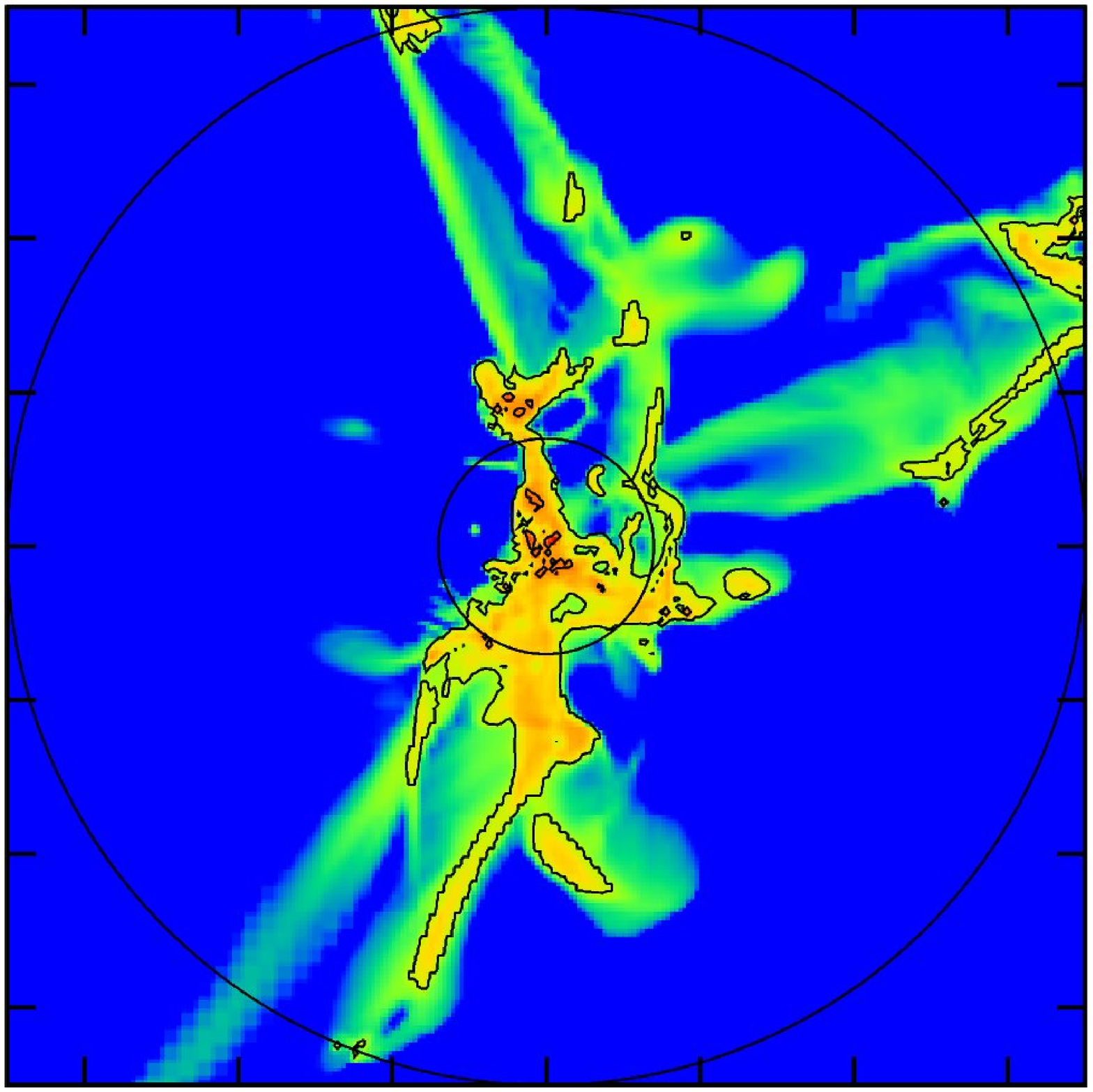}
\includegraphics[width=9.27cm]{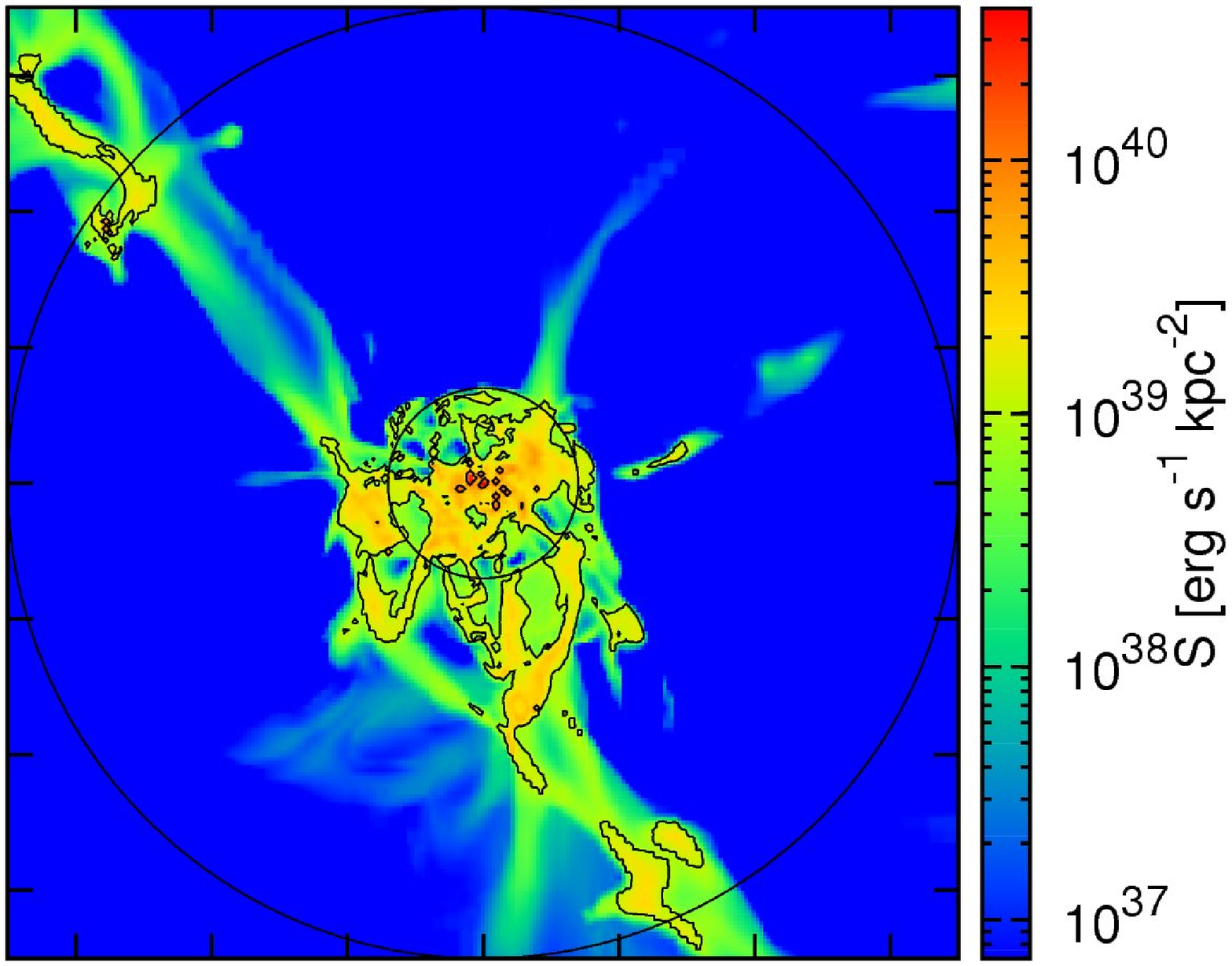}\\
\includegraphics[width=7.07cm]{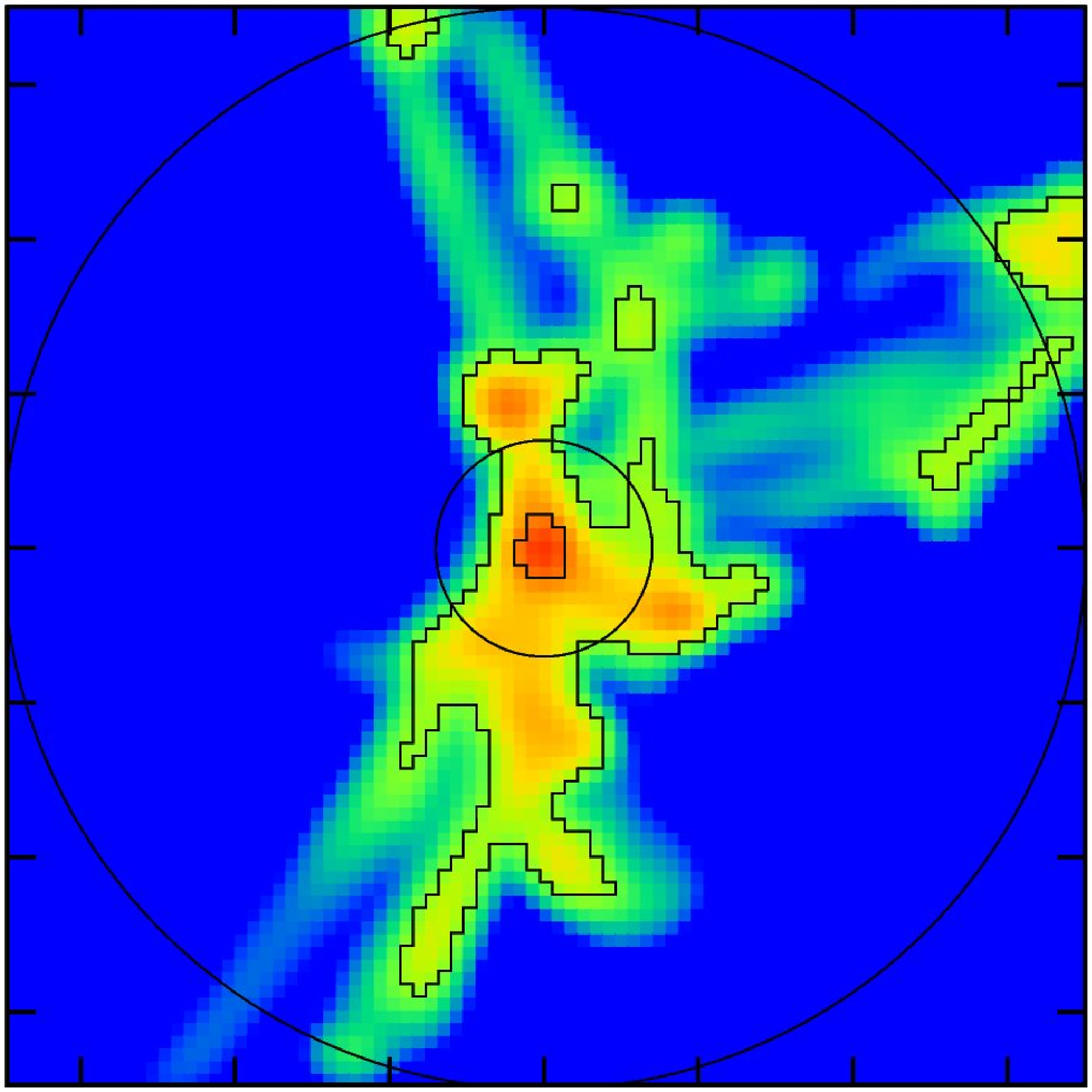}
\includegraphics[width=9.27cm]{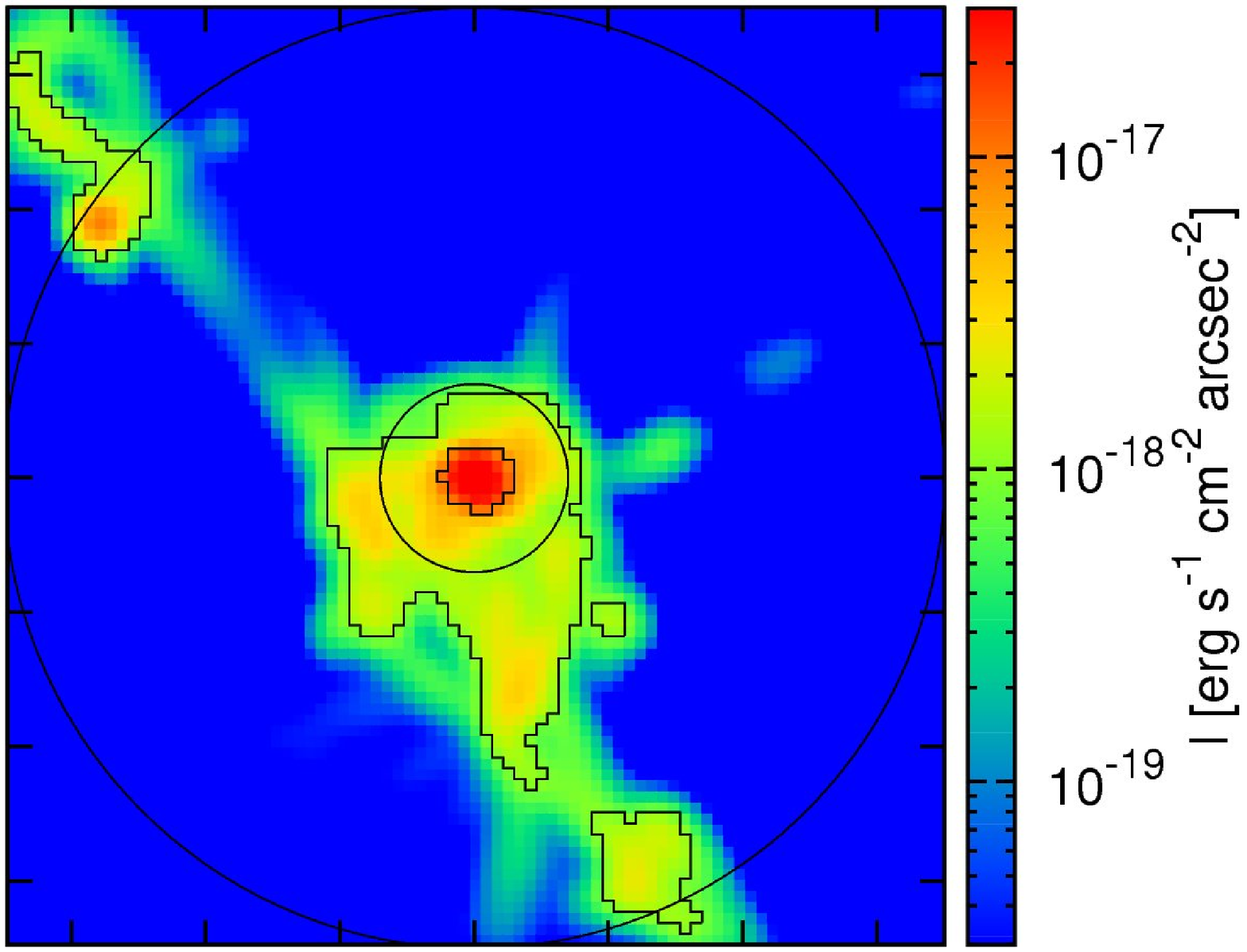}
\end{center}
\caption{Images of two simulated {\CD} galaxies at $z = 2.3$ (left and right).
The virial masses are $\Mv \simeq 4\times 10^{11}\msun$. The box side is
$140\kpc$ (physical). The outer circle marks the virial radius and the inner
circle is at $0.2\Rv$. Top: Neutral-hydrogen column density. Contours are shown
for $10^{20}$ and $10^{21}\cmms$. Middle: Restframe surface brightness $S$, with
$\fa=\falph$, showing contours at $10^{39}$ and $10^{40} {\rm erg\, s^{-1}\,
kpc^{-2}}$. Bottom: ``Observed" surface brightness $I$, at an angular
resolution of $\sim0.1$ arcsec, with contours at $10^{-18}$ and $10^{-17} {\rm
erg\, s^{-1}\, cm^{-2}\, arcsec^{-2}}$, corresponding to the contours of $S$.
The fraction of luminosity that originates from within these contours is 80\%
and 20\%, respectively.}
\label{fig:dscmaps}
\end{figure*}

\begin{figure*}
\begin{center}
\includegraphics[width=7.07cm]{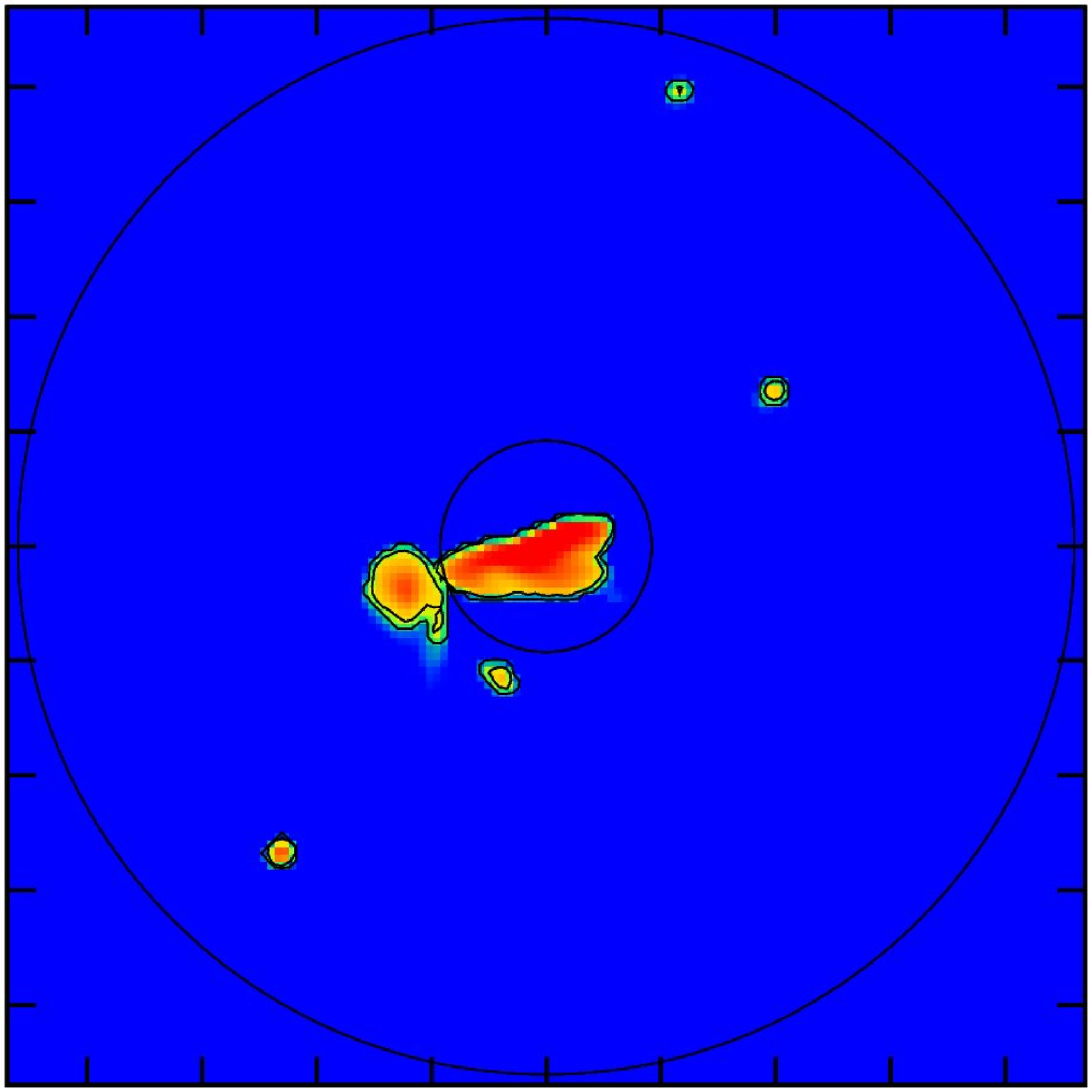}
\includegraphics[width=9.27cm]{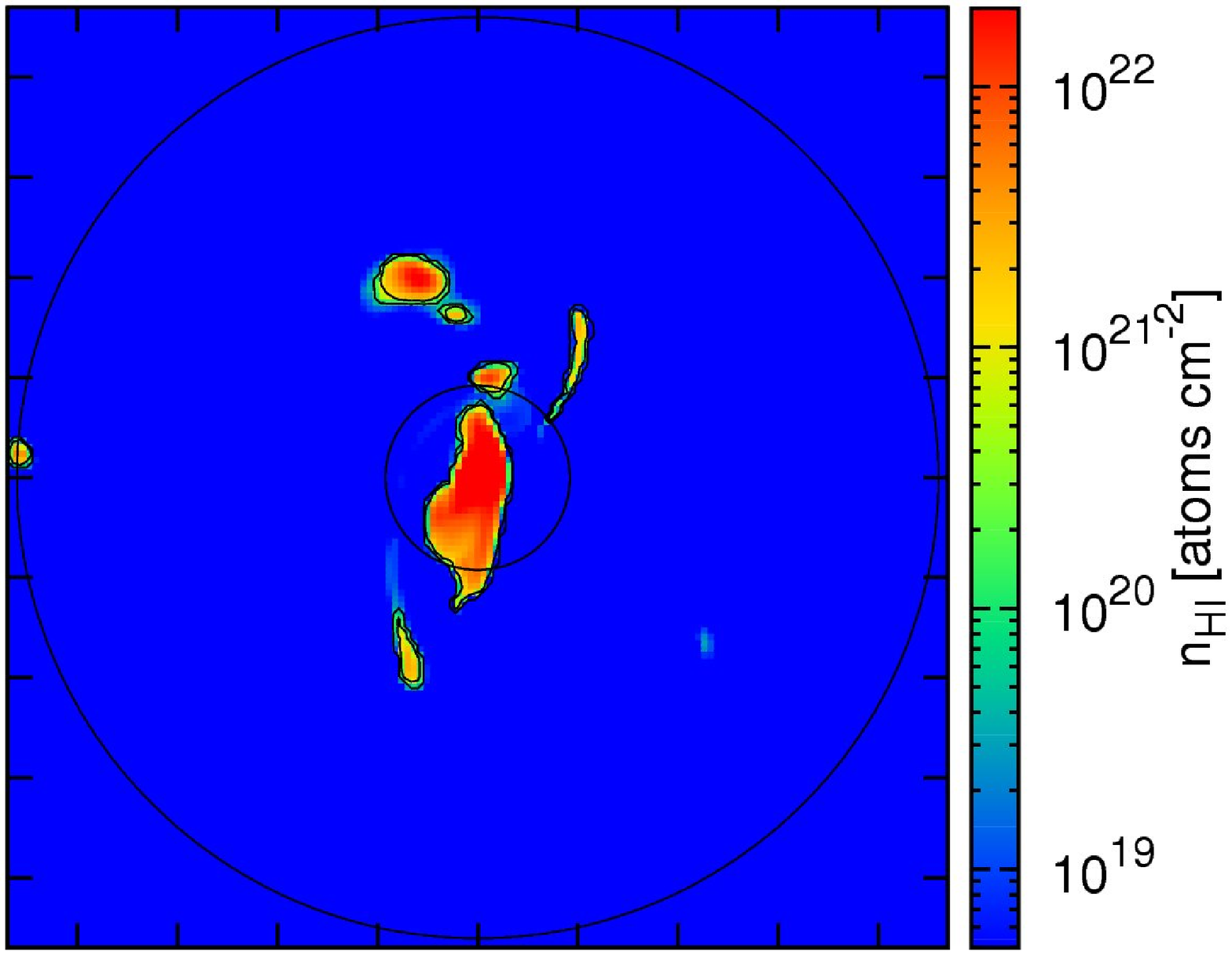}\\
\includegraphics[width=7.07cm]{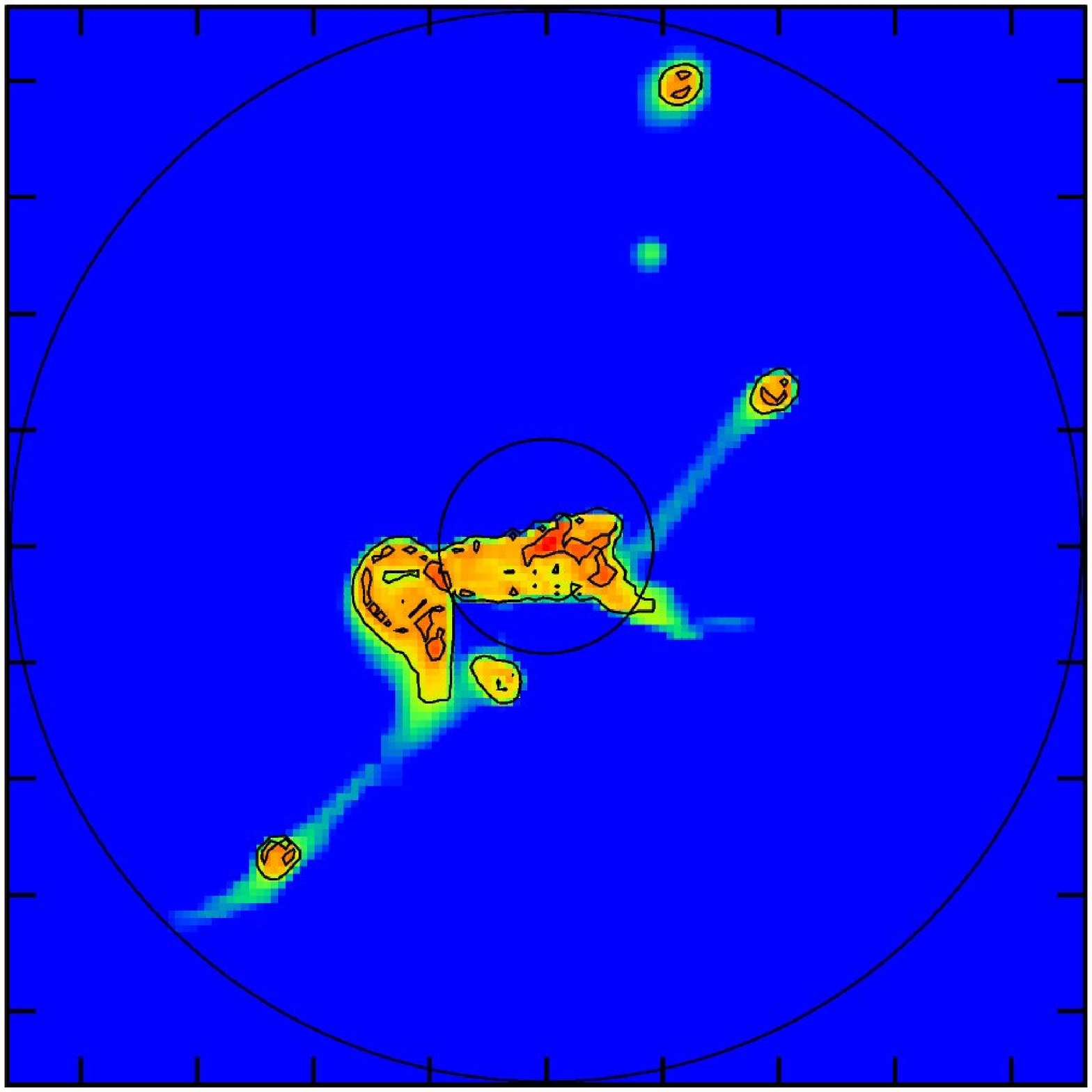}
\includegraphics[width=9.27cm]{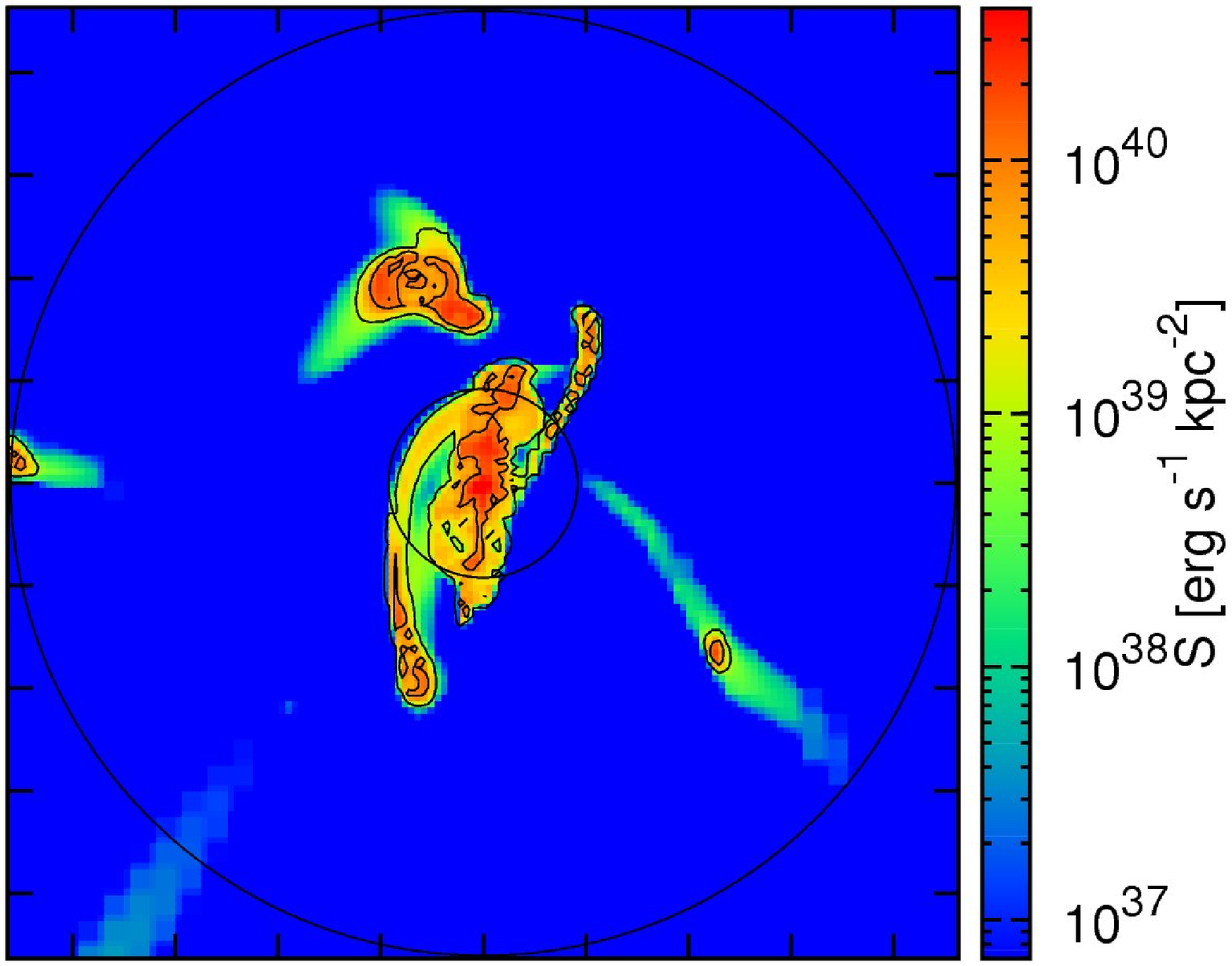}\\
\includegraphics[width=7.07cm]{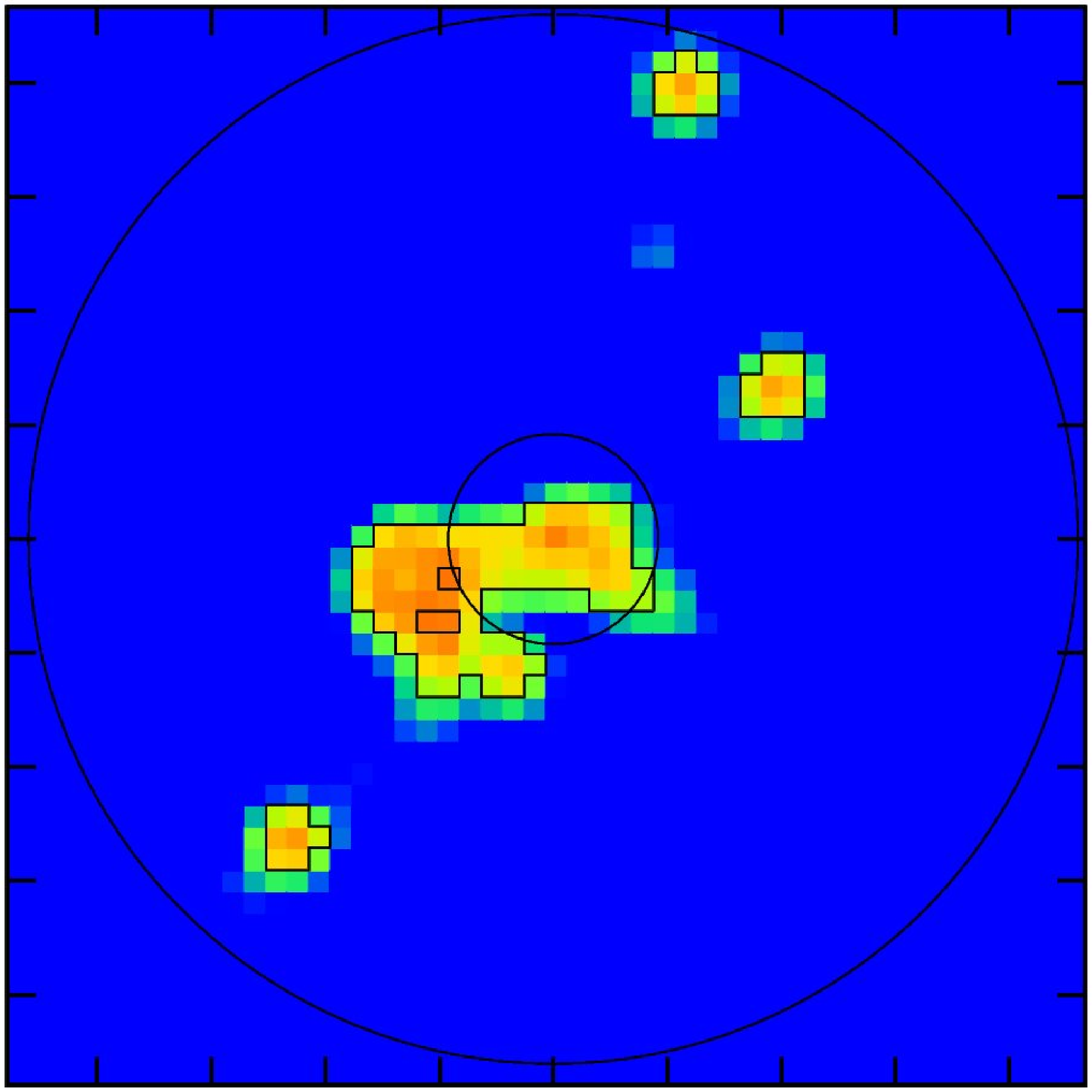}
\includegraphics[width=9.27cm]{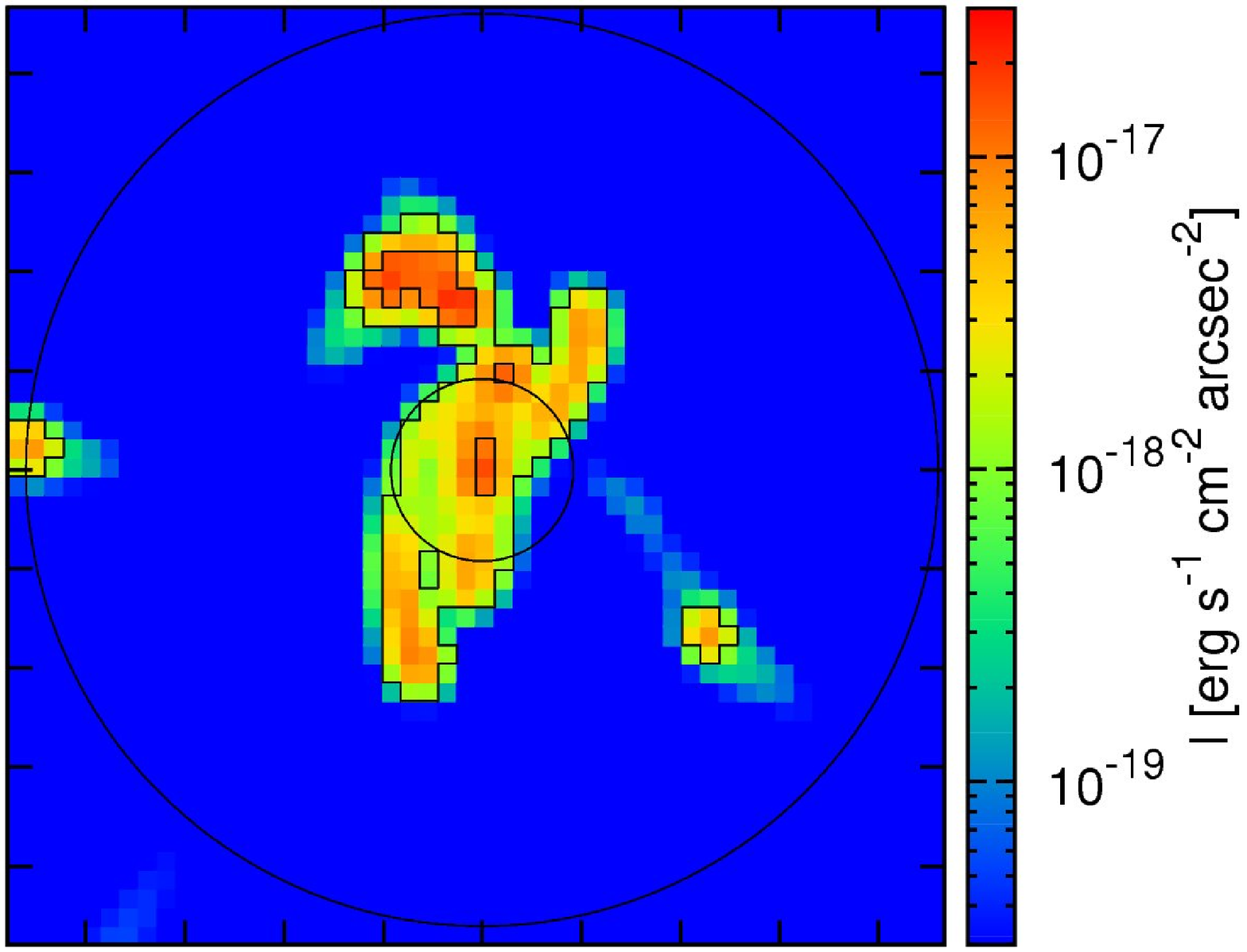}
\end{center}
\caption{Images of two simulated {\MN} galaxies at $z = 2.5$ (left and right).
The virial masses are $\Mv \simeq 10^{12}\msun$. The box side is $184\kpc$
(physical). The outer circle marks the virial radius and the inner circle is at
$0.2\Rv$. Top: Neutral-hydrogen column density. Contours are shown for
$10^{20}$ and $10^{21} \cmms$. Middle: Restframe surface brightness $S$, with
$\fa=\falph$, showing contours at $10^{39}$ and $10^{40} {\rm erg\, s^{-1}\,
kpc^{-2}}$. Bottom: ``Observed" surface brightness $I$, at an angular
resolution of $\sim0.1$ arcsec, with contours at $10^{-18}$ and $10^{-17} {\rm
erg\, s^{-1}\, cm^{-2}\, arcsec^{-2}}$, corresponding to the contours of $S$.
The fraction of luminosity that originates from within these contours is 93\%
and 18\%, respectively.}
\label{fig:mnmaps}
\end{figure*}

\Figs{dscmaps} and \ref{fig:mnmaps} show sample images of simulated galaxies,
two {\CD} galaxies and two {\MN} galaxies. The haloes are of virial mass
$\Mv \simeq 4\times 10^{11}$ and $10^{12}\msun$ respectively, and the
corresponding redshifts are $z=2.3$ and $2.5$.

The top panels present the neutral hydrogen column density as computed in
\se{lal} assuming CIE. The middle panels are maps of \lya restframe surface
brightness $S$, namely a fraction $\fa$ of the emissivity integrated along the
line of sight, as emitted per unit area at the galaxy. The bottom panels show
images of ``observed" surface brightness $I$ as an observer would see it, per
unit area in the galaxy and at the telescope. The restframe surface brightness
is converted to observed surface brightness via
\be
I = {S \over 4 \pi (1 + z)^4}
\, ,
\ee
In order to obtain realistic images, we applied a Gaussian PSF with a 0.6 
arcsec FWHM to mimic atmospheric distortions in good seeing conditions, and
assumed a pixel size of 0.2 arcsec at the telescope.

\begin{figure*}
\begin{center}
\includegraphics[width=6.05cm]{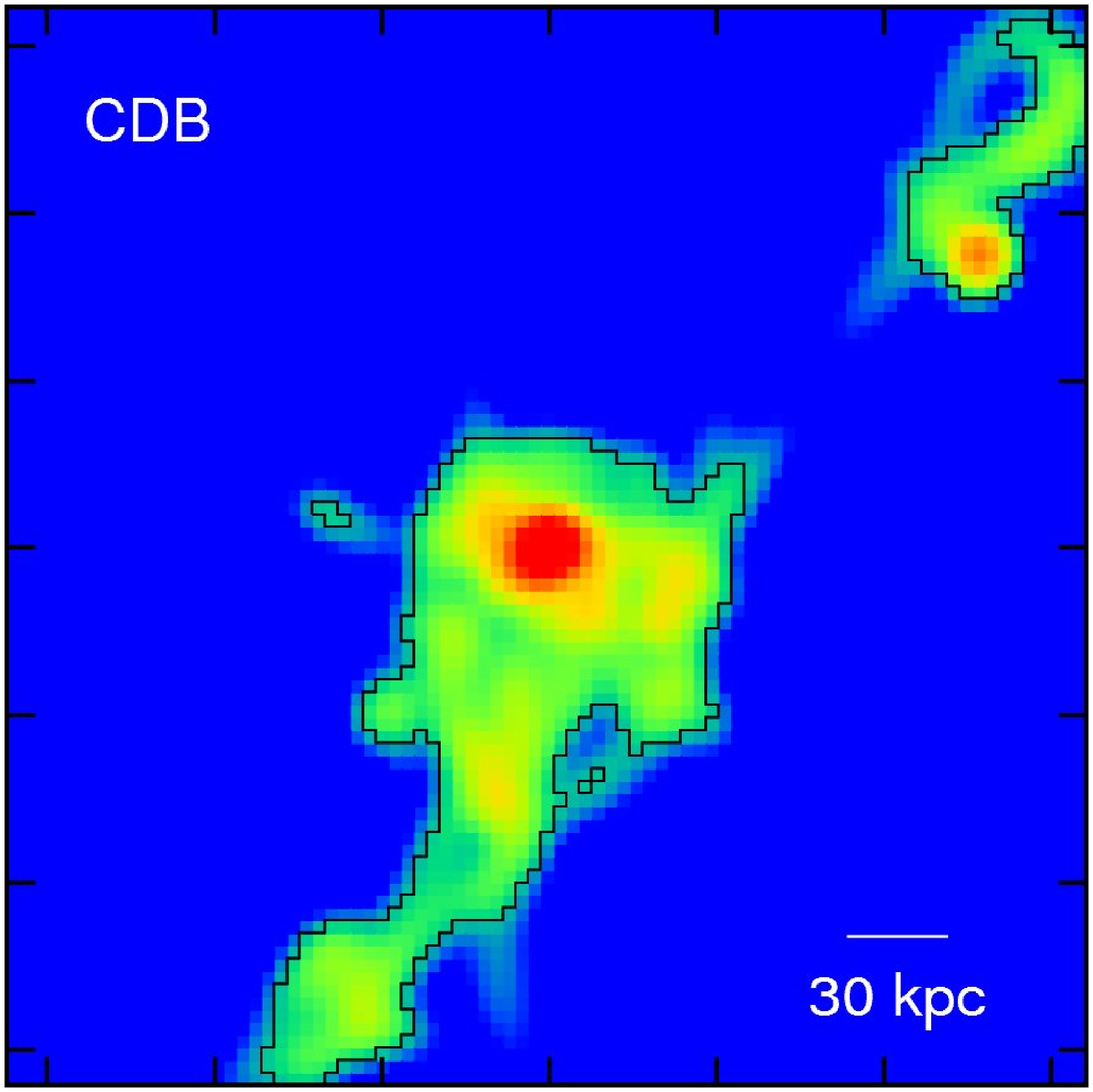}
\includegraphics[width=8.03cm]{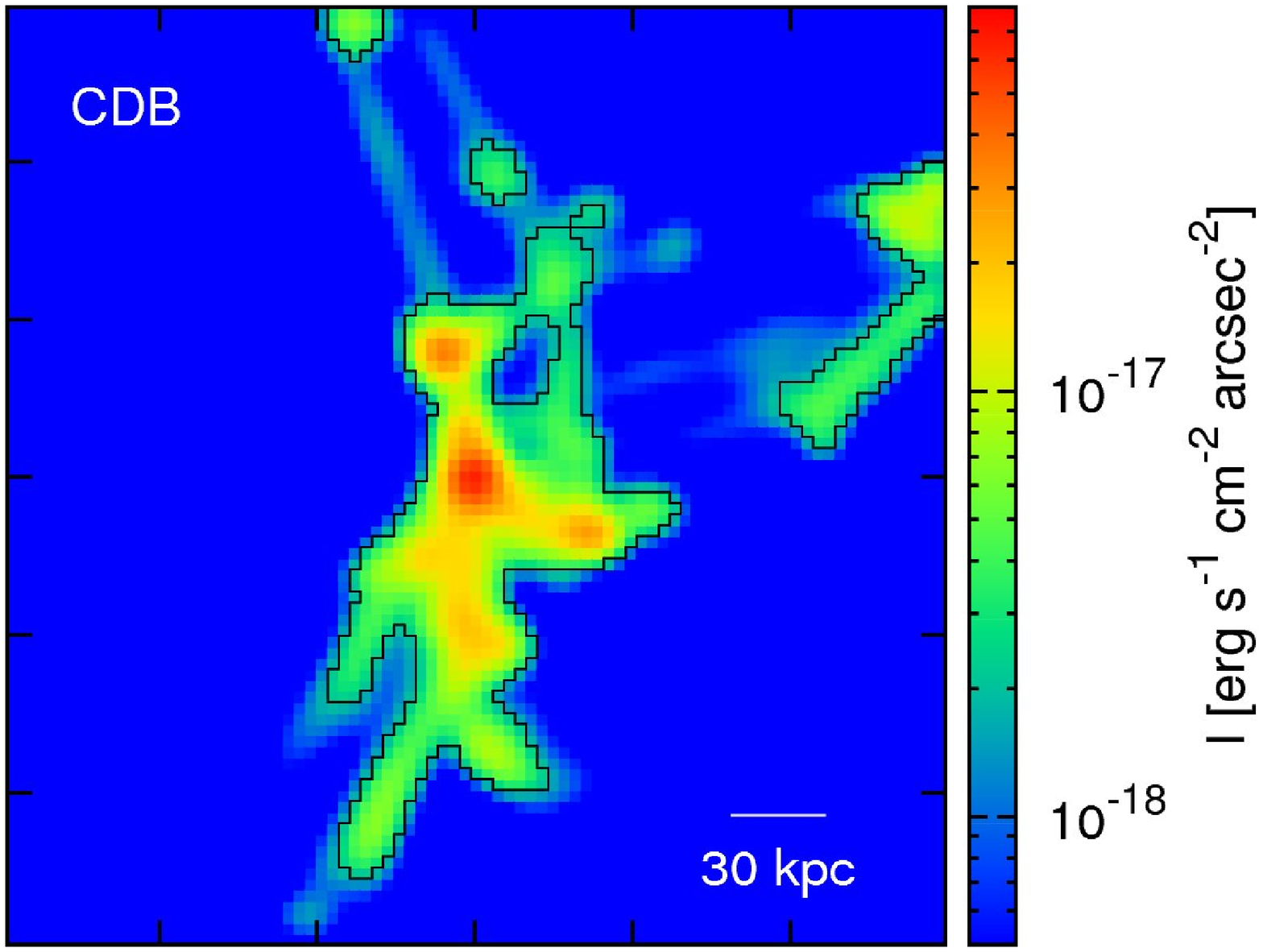}
\includegraphics[width=6.05cm]{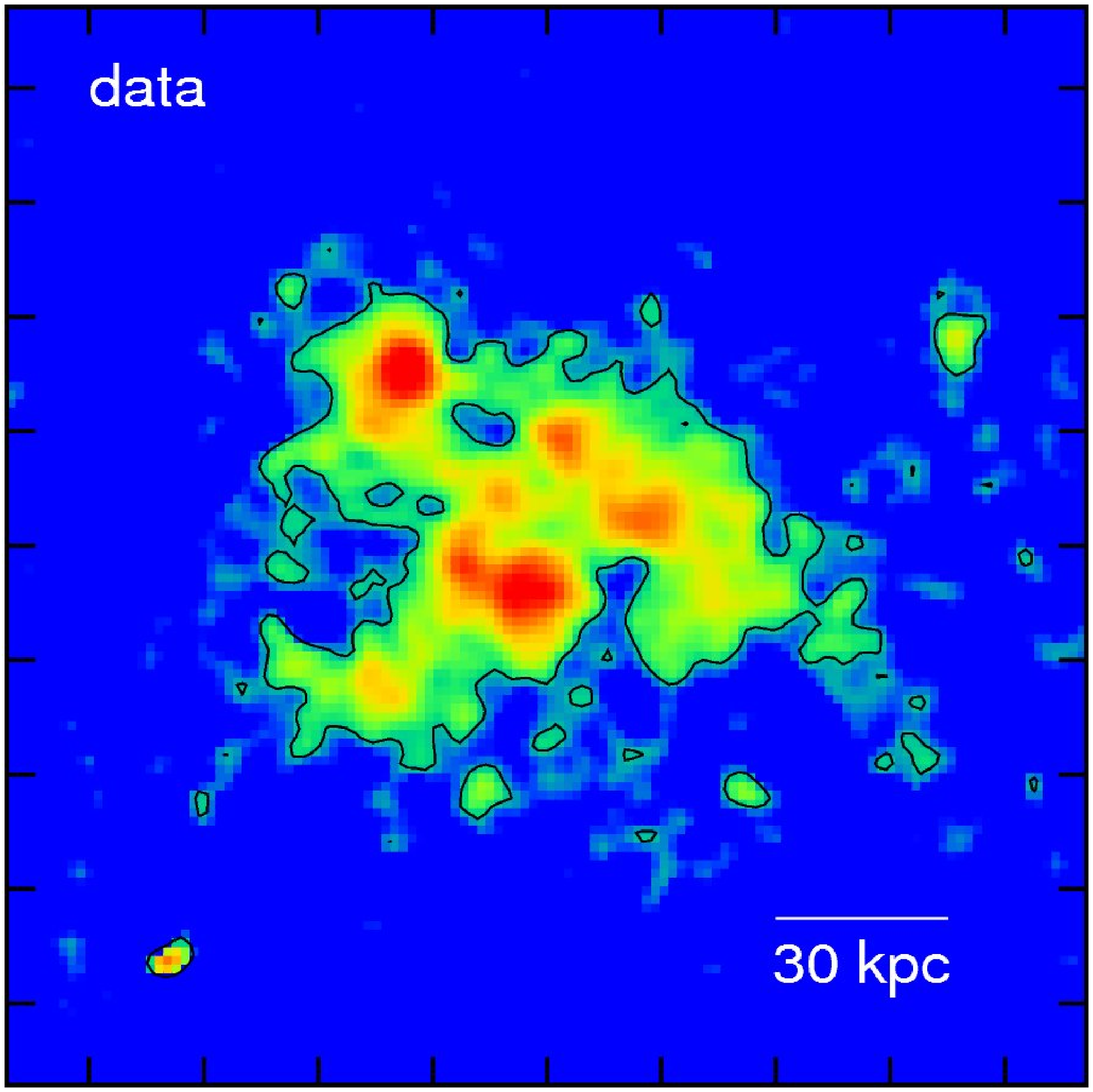}
\includegraphics[width=8.03cm]{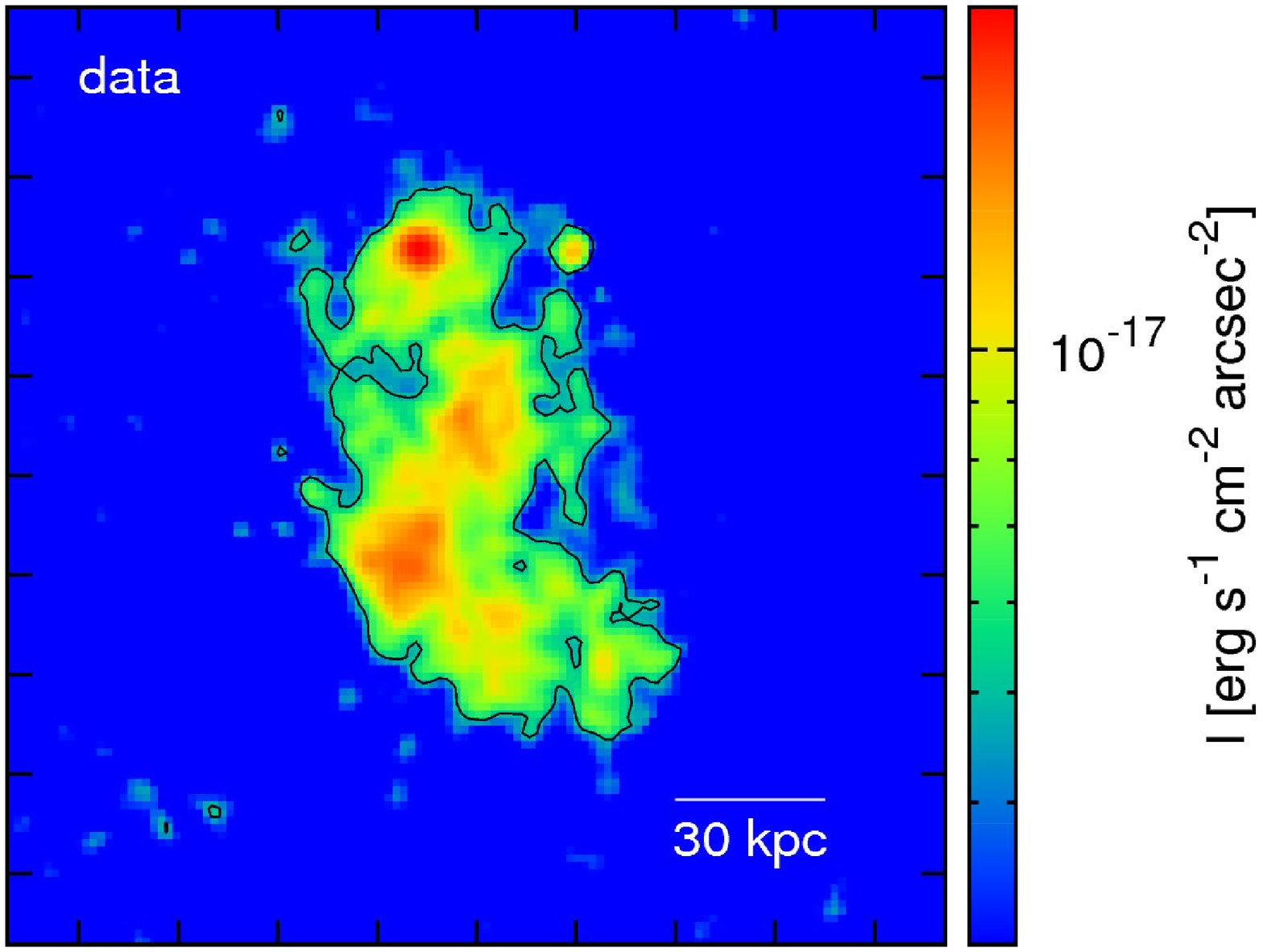}
\end{center}
\caption{A qualitative crude comparison of images of simulated (top) and
observed (bottom) LABs. The observed LABs are the two most luminous LABs from
\citet{matsuda} with $L \simeq 10^{44} \ergs$. The simulated images are of \CD
galaxies scaled from $z = 2.3$ to $z=3.1$ and to the luminosity of the
corresponding observed LABs using $I \propto R \propto L^{1/3}$ (the original,
unscaled luminosity was $5 \times 10^{43}$ erg s$^{-1}$). All four maps have an
angular resolution of $\sim 0.2$ arsec.}
\label{fig:mat}
\end{figure*}

One can read from the middle maps the fraction of the luminosity that comes
from each of the different parts of the halo. In particular, we consider
separately (a) the contribution from the disc galaxy and its near vicinity
inside a circle of radius $0.2\Rv$ and (b) the contribution from the halo at
$0.2-1.0\Rv$. We see in the figure that the restframe surface brightness in the
inner galaxy reaches values higher than $S \sim 10^{40}\, {\rm erg\, s^{-1}\,
kpc^{-2}}$ but over a limited area of $\sim 10^2\, {\rm kpc^2}$, thus
contributing $L \sim $ a few $\times 10^{42}$ erg s$^{-1}$. The typical surface
brightnesses in the halo are lower by a factor of a few, but the emission
regions are spread over an area that is larger by an order of magnitude, thus
providing a comparable contribution to the total luminosity. Most of the
luminosity comes from the low-density streams, and a non-negligible fraction is
from clumps associated with the streams.

Recall that the emission from the disc may be partly absorbed by dust, which is
ignored here, but we do not expect substantial dust absorption in the streams in
the outer halo, where the metallicity is 0.01 to 0.1 solar. We can thus
consider the halo contribution to be a safe lower limit to the overall \lya
luminosity, and take the luminosity as estimated from the whole system of disc
plus halo as an upper limit.

In the bottom panels we see blobs with surface brightness above $10^{-18}
\ergsc$ extending to $\sim 100$ kpc. The images are clumpy and irregular. Their
shapes are non-circular; they tend to be elongated and asymmetric, showing
finger-like extensions. These sample images resemble the observed images of
LABs, e.g., those shown in Figure~8 of \citet{matsuda}. \Fig{mat} compares
``observed" surface brightness maps of two simulated \CD galaxies (from the
bottom panels of Figures \ref{fig:dscmaps} and \ref{fig:mnmaps}) with the
images of the two most luminous LABs observed by \citet{matsuda}. The
qualitative similarity of the irregular, clumpy and elongated LAB morphologies
and their sizes is encouraging.

\section{Lyman-alpha Luminosity as a function of mass and redshift}
\label{sec:L(M,z)}

\begin{figure*}
\begin{center}
\includegraphics[width=8.69cm]{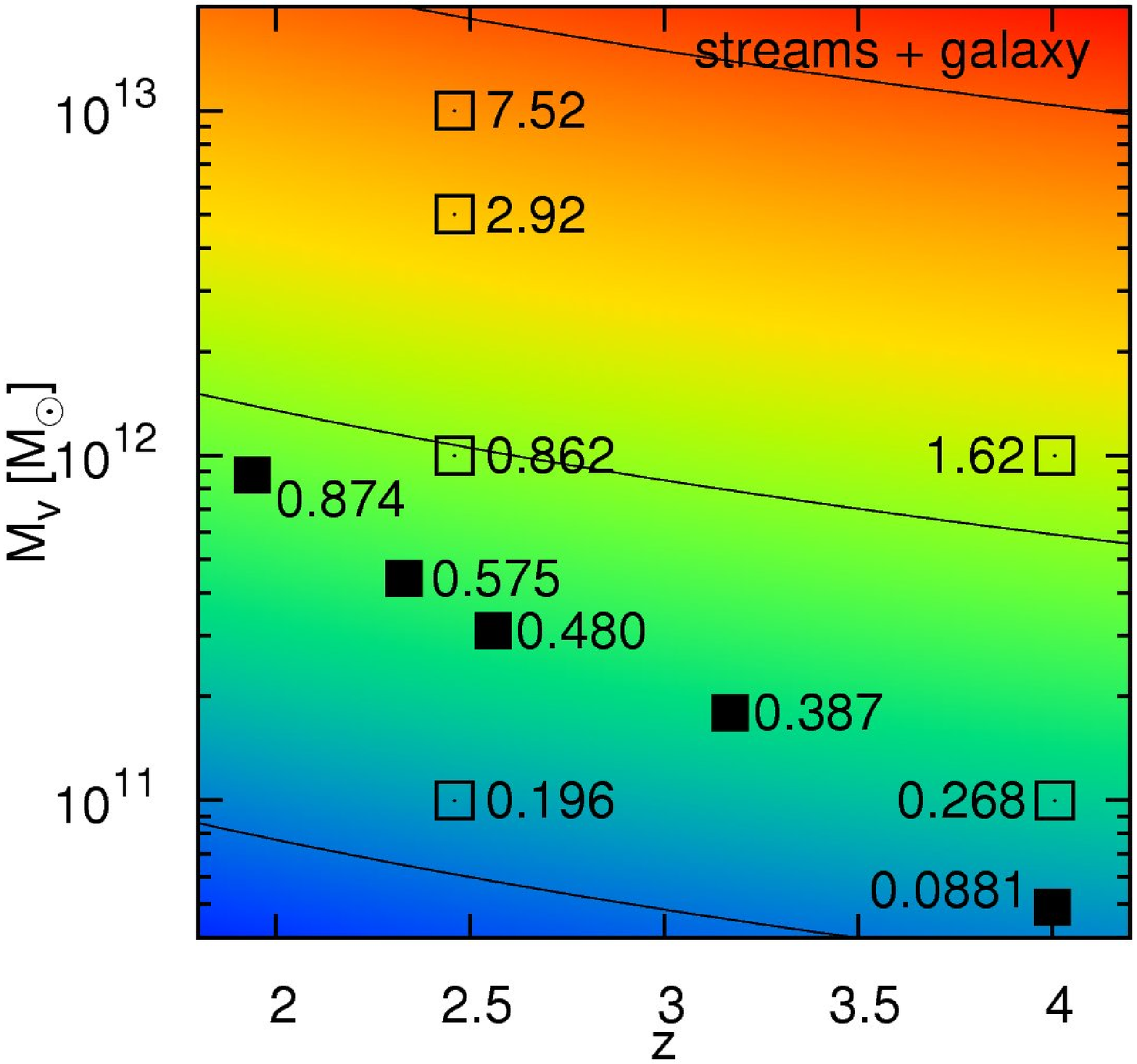}
\includegraphics[width=8.91cm]{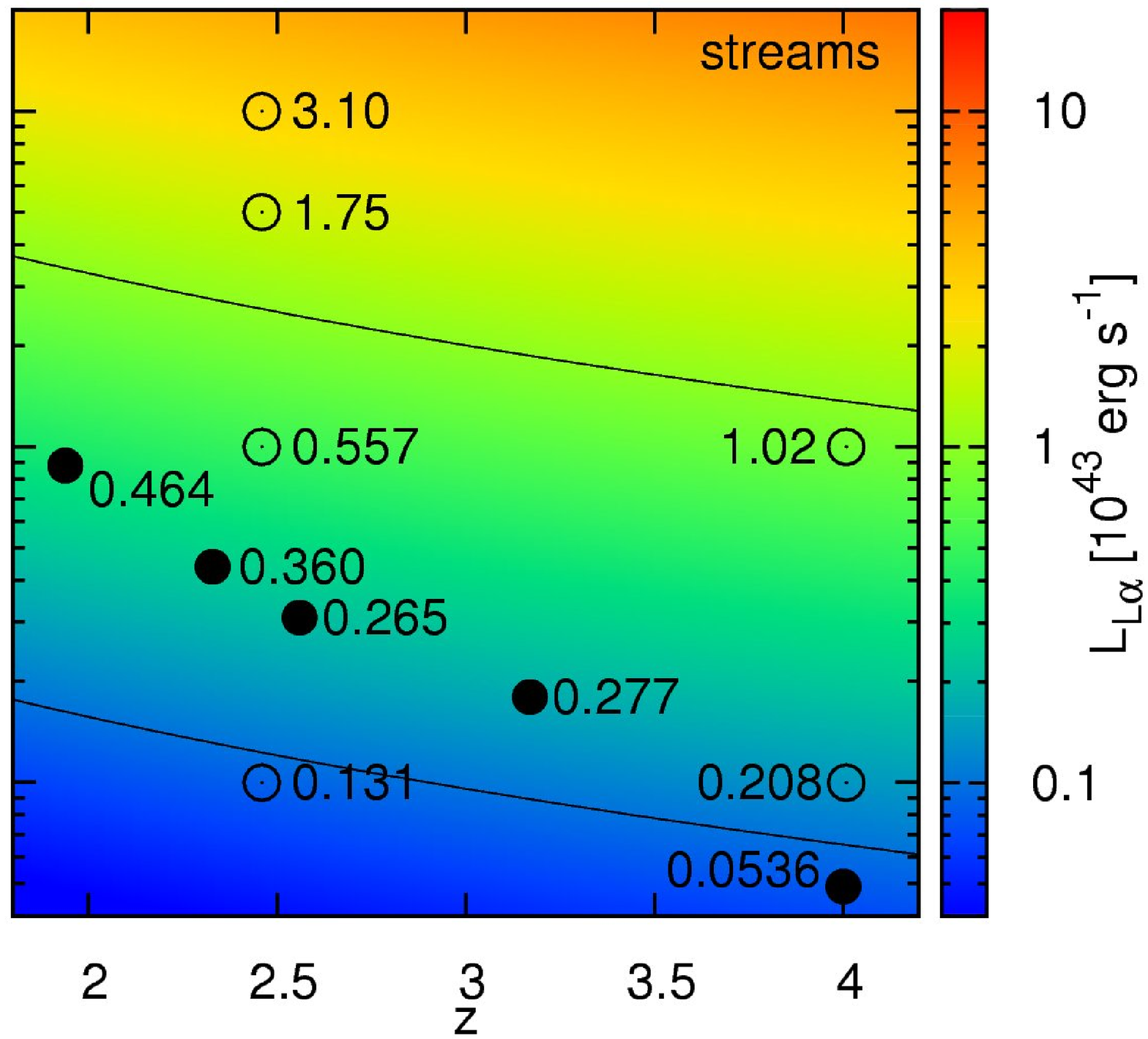}
\end{center}
\caption{Total \lya luminosity as a function of halo mass and redshift,
separately for streams+galaxy ({\it left}) and for streams only ({\it right}).
The average values as drawn from the simulations in bins of $(\Mv,z)$ are
marked by symbols and numbers in units of $10^{43}\ergs$. The \CD results are
marked by filled symbols and the \MN results by open symbols. Shown in colour
is the fitting function \equ{lm} with contours at 10$^{42}$, 10$^{43}$ and
10$^{44}$ erg s$^{-1}$.}
\label{fig:mzco}
\end{figure*}

The variation of average total \lya luminosity from within the virial sphere as
a function of halo mass and redshift, as derived from the simulations, is shown
in \fig{mzco}. The two panels refer to the luminosity from the whole halo,
streams and central galaxy, as an upper limit, and to the streams at $r>0.2\Rv$
only as a lower limit. The fraction of the radiation emitted is set to
$\fa=\falph$, as determined by a fit to the observed luminosity function in
\se{lumfun} below. The luminosities are quoted for small bins about points in
the $(\Mv,z)$ plane. Four points refer to the average luminosities over the
three {\CD} galaxies at $z=1.9, 2.5, 3.2$ and 4.0. Each of the other points
refer to the average over 12 {\MN} galaxies. The two-dimensional functional fit
to the quoted values, $L(\Mv,z)$, is shown in colours. We use a function of the
form
\be
L_{43}(\Mv) = A\, \left(M_{12}\right)^B\, (1+z)^C \, ,
\label{eq:lm}
\ee
where $A$, $B$ and $C$ are free parameters, $L_{43}\equiv L_{\La}/10^{43}\ergs$
and $M_{12}\equiv \Mv/10^{12}\msun$. To determine the free parameters, we first
address the mass dependence at a fixed redshift where we have performed our
most detailed analysis in {\CD}, $z \simeq 2.4$. We then assume that the same
mass dependence is valid between $z\sim 2$ and $z \sim 4$, and evaluate the
redshift dependence. 

\begin{figure}
\begin{center}
\includegraphics[width=8.4cm]{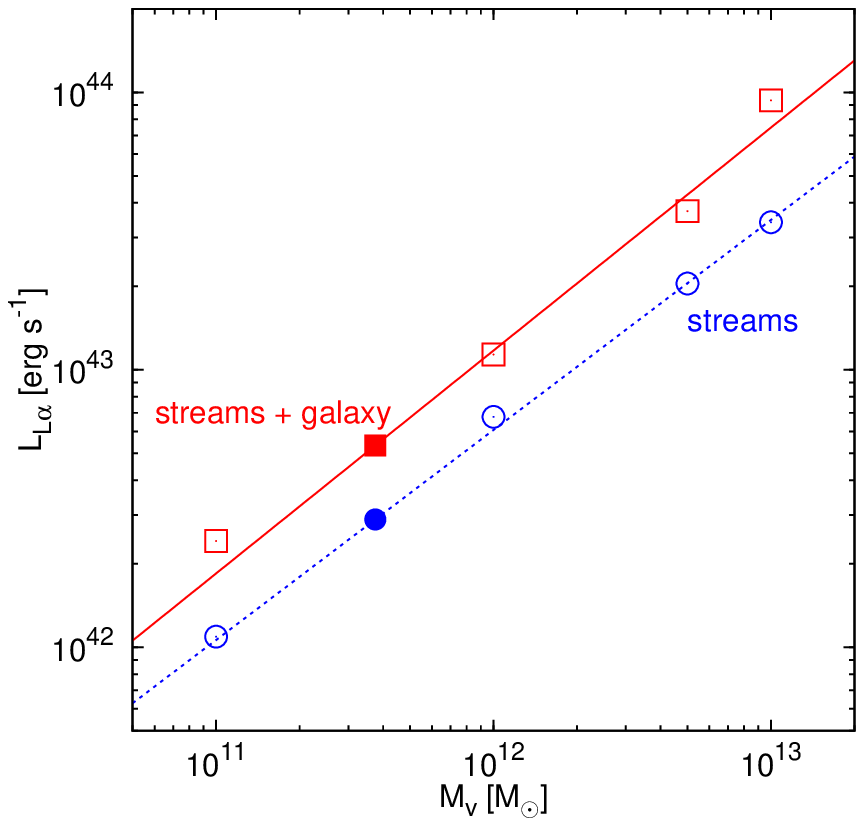}
\end{center}
\caption{\lya luminosity as a function of halo mass at $z = 3.1$ with
$\fa=\falph$, for the total luminosity within the virial radius (upper, red
symbols and curve), and for the halo outside $0.2\Rv$ (lower, blue symbols and
curve). Each symbol represents the average luminosity over simulated haloes of
a given mass, 12 haloes from the {\MN} simulation (open symbols) and 3 from the
{\CD} simulations (filled symbols). The lines are least-squares fits using
\equ{lm} with slopes $B=0.80$ and $0.76$, respectively.}
\label{fig:lumiz246}
\end{figure}

\Fig{lumiz246} shows the luminosity as a function of halo mass at $z = 3.1$.
The luminosity is computed within the isophotal surface brightness threshold of
$2.2 \times 10^{-18}$ erg s$^{-1}$ cm$^{-2}$ arcsec$^{-2}$, also used by
\citet{matsuda}. We adopted $\fa=\falph$. A small correction upward of $\sim
30\%$ has been applied to the results as extracted at $z \simeq 2.4$ in order
to bring them to $z=3.1$ in this plot, using \equ{lm} with $C=1.3$ as
determined below. The average luminosity for the three {\CD} galaxies is shown
at $\Mv \simeq 4\times10^{11} \msun$. Each of the other symbols is the average
of 12 {\MN} galaxies. We see that in the mass range most relevant for LABs,
$\Mv = 10^{11.5}-10^{12.5} \msun$, the luminosity is roughly proportional to
halo mass. This dependence is driven by the near-linear mass dependence of the
accretion rate, \equ{acc}.
 
For a more accurate description of the mass dependence we fit the function of
\equ{lm} to the results at the fixed redshift $z \simeq 2.4$, using
least-squares fit in the log, and obtain the lines shown in \fig{lumiz246}
with $B = 0.80$ for the whole halo and $B=0.76$ for the streams only.

\begin{figure}
\begin{center}
\includegraphics[width=8.4cm]{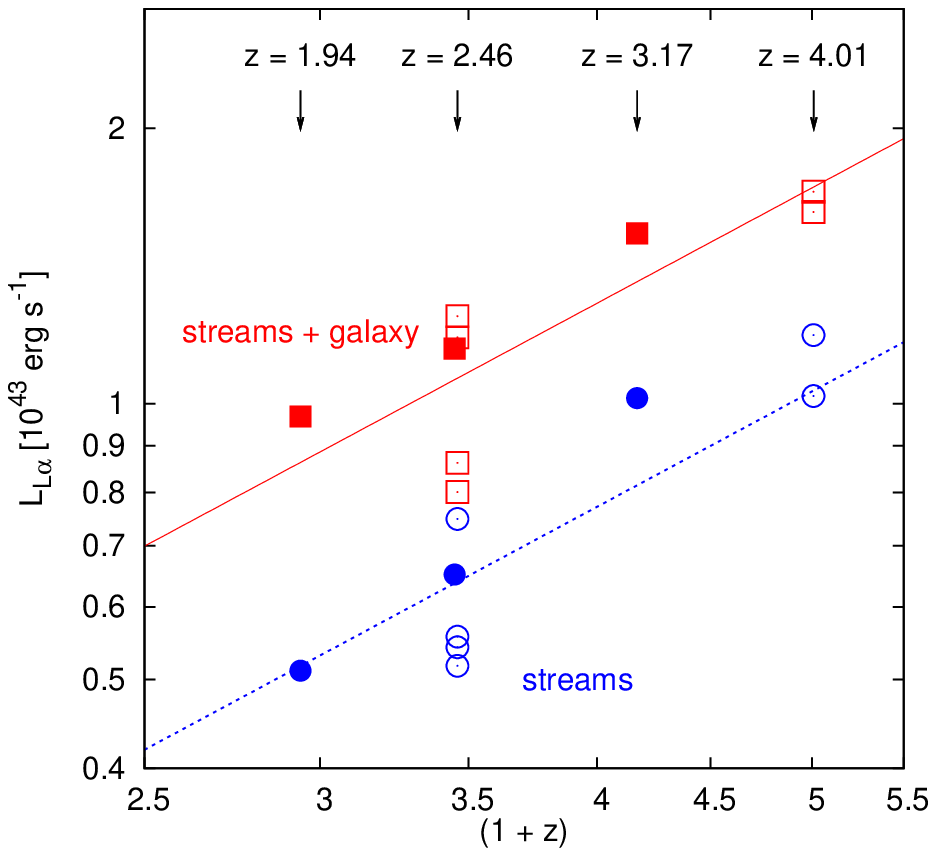}
\end{center}
\caption{\lya luminosity as a function redshift for $\Mv = 10^{12}\msun$ with
$\fa=\falph$, for the total luminosity within the virial radius (upper, red
symbols and curve), and for the halo outside $0.2\Rv$ (lower, blue symbols and
curve). Each symbol represents the average luminosity over simulated haloes of
a given mass, 12 haloes from the {\MN} simulation (filled symbols) and 3 from
the {\CD} simulations (open symbols). The luminosities were scaled from the
simulated values at the different halo masses to $\Mv = 10^{12}\Msun$ using the
mass dependence obtained in \equ{lm}. The curves are approximate eye-ball
power-law fits with a slope $C=1.3$.}
\label{fig:lumiallz5}
\end{figure}

We next use this functional mass dependence to scale the simulated results at
other redshifts to a fixed mass, $\Mv=10^{12}\msun$. The results are the values
as quoted in the $(\Mv,z)$ bins shown in \fig{mzco}. The scaled luminosities
are shown in \fig{lumiallz5}. A crude power-law fit to the symbols yields a
redshift-dependence power index $C=1.3$, with the normalisation parameters
$A=0.188$  and $0.0972$ for streams+galaxy and for streams only, respectively.
We note that the obtained redshift dependence of the luminosity is somewhat
weaker than the $(1+z)^{2.2}$ dependence of the accretion rate in \equ{acc}.
The accuracy of this scaling with redshift is not an important issue for us
here since we only use it to correct the luminosities from $z \simeq 2.4$ to
$z=3.1$, a correction of less than 30\%.

\section{Lyman-alpha Luminosity Function}
\label{sec:lumfun}

\begin{figure}
\begin{center}
\includegraphics[width=8.4cm]{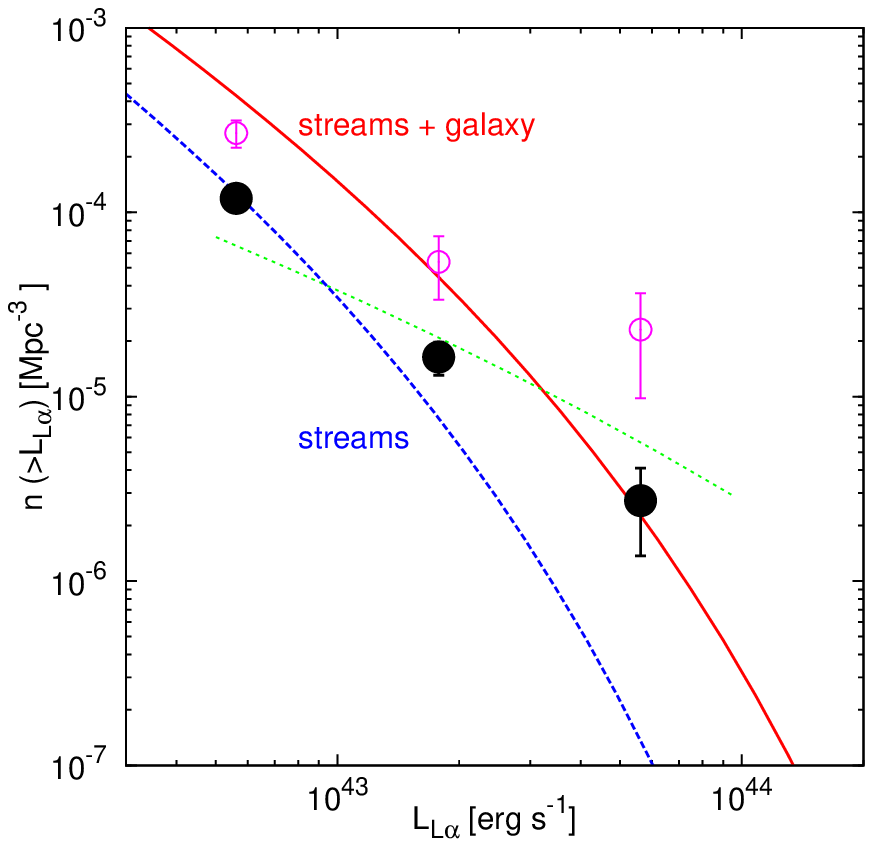}
\end{center}
\caption{\lya luminosity function, showing the comoving number density of LABs
brighter than $L_{\La}$ at $z=3.1$. The observed symbols (black, filled circles)
are based on a sample of 201 LABs by Matsuda (private communication) in a fair
sample. The earlier, higher estimates (magenta, open circles) are based on 35
LABs in a dense cluster environment. The lower and upper limits for the
luminosity function as derived from the simulations are shown, for streams only
(dashed, blue curve) and for streams+galaxy (solid, red curve), respectively.
The transmission factor was set to $\fa=\falph$ for a best match between theory
and observation. The gravitational-heating toy-model prediction based on
\equ{eheat} is shown (green, dashed line).}
\label{fig:st}
\end{figure}

To obtain a predicted LAB luminosity function at $z \sim 3$, we convolve
$L_{\La}(\Mv)$ with the Sheth-Tormen halo mass function \citep{st} at the same
redshift. Here we assume a cosmology with $\Omega_\Lambda = 0.72$,
$\Omega_{\rm M} = 0.28$, $\sigma_8 = 0.83$, $h = 0.71$ and $\Omega_{\rm B} =
0.045$. The resultant luminosity function is presented in \fig{st}, both for
the total luminosity within the virial radius and for the halo streams only, in
the range $0.2-1.0 \Rv$, which we consider to be upper and lower limits
respectively given the uncertainty in the dust obscuration from the inner
galaxy. The fraction of emitted radiation, which could range in principle
between 0.5 and 1.0, is set to $\fa=\falph$.  

We computed a crude estimate of the observed \lya luminosity function at $z
\sim 3.1$ in 3 luminosity bins using the most recent 201 LABs observed by
Matsuda et al. (private communication). This sample contains three fields of
different environment densities, and can thus be crudely considered as a fair 
sample. The number densities were determined by counting the number of blobs in
the luminosity bins and dividing by the comoving survey volume of $V_{\rm s} =
1.5 \times 10^6$ Mpc$^3$ (or physical volume $3.6 \times 10^5$ Mpc$^3$). Also
shown is the luminosity function derived earlier from a partial sample of 35
LABs in a dense cluster environment \citep{matsuda}\footnote{The comoving survey
volume here is $V_{\rm s} = 1.3 \times 10^5$ Mpc$^3$ (or physical volume $3.2
\times 10^4$ Mpc$^3$). They used a cosmology with $\Omega_{\rm M} = 0.3$,
$\Omega_\Lambda = 0.7$ and $h = 0.7$, which is very close to the cosmologies
used in the simulations and in the analysis. They choose their sample as
explained in their section 3. The observed luminosity function was computed for
a combined limit of surface brightness $(2.2 \times 10^{-18}$ erg s$^{-1}$
arcsec$^{-2})$ and size (16 arcsec$^{-2}$), while in the simulations we use
only a surface brightness limit. Had we imposed a similar size limit in the
simulations, one can deduce from \fig{area} that our predicted luminosity
function would have been suppressed at luminosities below a few times
$10^{42}\ergs$, making the comparison with the counts at the faint end less
certain.}. The latter overestimates the universal luminosity function by a
factor of a few as expected from the high environment density. A Poissonian
error has been attached to every bin. The number density has been derived by
dividing the number of objects in each bin by the total survey volume. 

We learn from \fig{st} that the predicted luminosity function has a similar
slope to the observed one. For $f_{\alpha} \sim \falph$, the predicted upper
and lower limits border the observed function from above and below throughout
the whole range of LAB luminosities. We find that $f_\alpha$ changes roughly
linearly with the adopted cosmological $\sigma_8$ through its effect on $n(M)$.
Therefore, the $\sim 10\%$ uncertainty in $\sigma_8$ translates into a similar
uncertainty in $f_\alpha$.

Also shown is the luminosity function as estimated from our crude toy model for
gravitational heating, \se{feasibility}, again convolved with the Sheth-Tormen
halo mass function, and using $\fa \fc =\fcfa$. It is remarkable that this very
crude toy model recovers the simulation results in the relevant mass range to
within a factor of a few.

\section{Surface-Brightness Profile, Area and Linewidth}
\label{sec:obs}

A detailed comparison of theory to data will be presented in a future paper
where we apply a more accurate radiative transfer calculation to the simulated
galaxies. However, it is worthwhile to report here on a preliminary comparison
with data, based on the simplified analysis of the current paper, which is 
encouraging.

\subsection{Surface-brightness profile}
\label{sec:profile}

\begin{figure}
\begin{center}
\includegraphics[width=8.4cm]{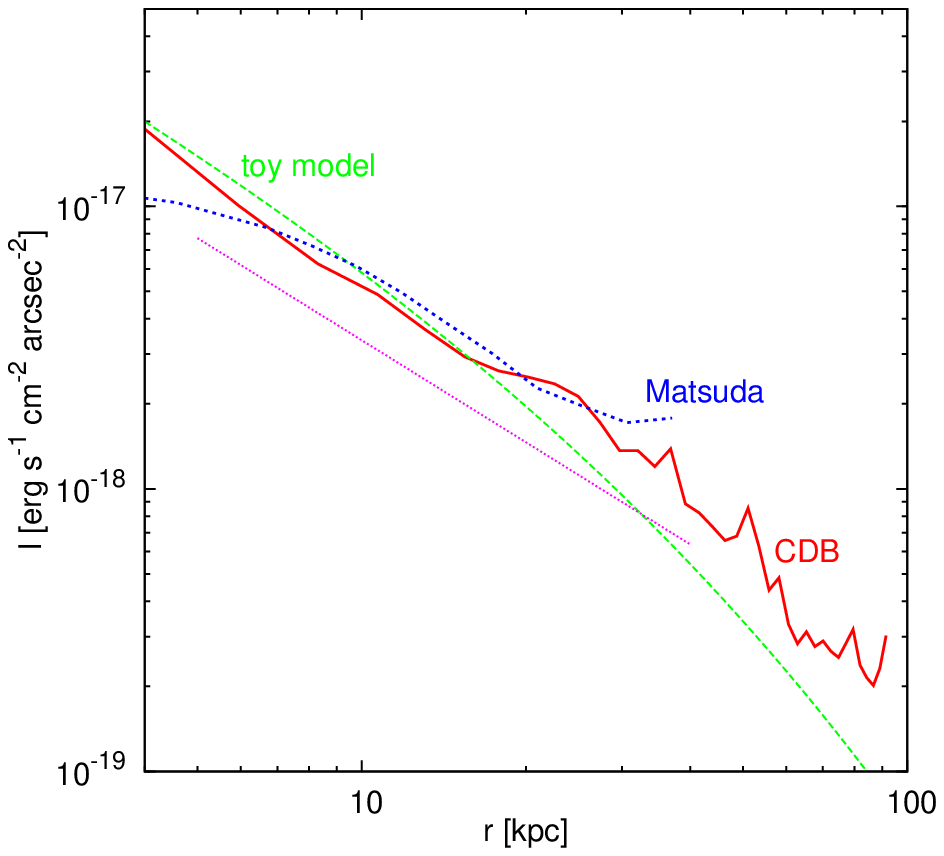}
\end{center}
\caption{\lya surface-brightness profile. The stacked profile from the three
{\CD} galaxies (red, solid line) scaled from $z = 2.3$ to $z = 3.1$ is compared
to the stacked profile of 35 observed LABs from \citet{matsuda} at $z=3.1$
(blue, dotted curve). The individual profiles are scaled to a LAB of a fiducial
luminosity of 10$^{43}$ erg s$^{-1}$ (see text). The toy-model prediction from
section \ref{sec:feasibility} is shown (green, dashed curve). A power law of
slope $-1.2$ is shown as a reference (pink, dotted line). There is a good
agreement between simulation and data in the radius range $10-30$ kpc, where
the observed profile is meaningful.}
\label{fig:surfbright}
\end{figure}

\Fig{surfbright} shows the surface-brightness profile of the stacked images
from the three \CD simulations, assuming $\fa=\falph$ throughout the whole
halo. It is compared to stacked profiles of 35 observed LABs from
\citet{matsuda}. For the stacking of the observed LABs, which range over more
than an order of magnitude in luminosity, we scale the LAB radius $R$ and
surface brightness $I$ to a fiducial luminosity $L_0 = 10^{43}$ erg s$^{-1}$.
We write $L \propto I\, R^2$ and assume $L \propto \Mv$ and $R \propto \Rv
\propto \Mv^{1/3}$ to obtain the scaling $R \propto I \propto L^{1/3}$. Thus,
for a LAB with luminosity $L$, we multiply both the radius and the
surface-brightness by the factor $(L_0/L)^{1/3}$. Non-zero values for $I$ are
considered for each LAB only above the isophotal surface-brightness threshold
of $2.2 \times 10^{-18}$ erg s$^{-1}$ cm$^{-2}$ arsec$^{-2}$ and only the inner
adjacent non-zero values (i.e. all values outside the innermost zero value are
also treated as zero values) were used in the averaging of $I$ at each radius
$r$. The stacked simulated profile has been scaled accordingly to match the same
fiducial luminosity $L_0$.

We see that a power law $I(r) \propto r^{-\gamma}$ is a good fit to the
simulated profile, with $\gamma \simeq 1.2$, from the disk scale $r< 10$ kpc
out to $\sim 40$ kpc, which is about half the virial radius. It seems to
steepen to $\gamma \sim 2$ at larger radii.\footnote{Note that the slope
$\gamma \simeq 1.2$ is consistent with the slope of the cumulative 3D
luminosity profile shown in \fig{gravpot}, where $L(<r) \propto r^{0.8}$ in the
range $r= (0.1-0.4)\,\Rv$.} The observed images of \citet{matsuda} are subject
to a point spread function of $\sim 1$ arcsec, which is responsible for the
flattening of the observed profile at $r \leq 10$ kpc. Given the high
background outside the surface-brightness threshold, the meaningful part of the
observed stacked profile is limited to the range $10-30$ kpc. In this range,
the power law with a slope $\gamma = 1.2$ provides a good fit to the observed
profile as well. We also see that the crude toy model of \se{feasibility}
provides a profile that is consistent with the simulated profile to within a
factor of two, as seen in \Fig{gravpot}.

\subsection{Isophotal area}

Two of the observable global quantities for each LAB are the area encompassed 
by an isophotal contour of a given threshold surface brightness, and the
corresponding luminosity. In order to compare to the data of \citet{matsuda},
we applied to our simulated galaxies, scaled to $z=3.1$, a threshold of $2.2
\times 10^{-18}$ erg s$^{-1}$ cm$^{-2}$ arcsec$^{-2}$.

\begin{figure}
\begin{center}
\includegraphics[width=8.4cm]{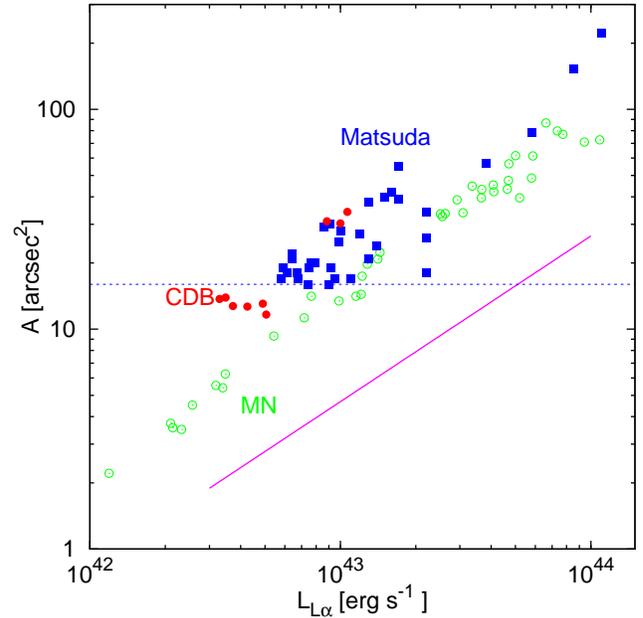}
\end{center}
\caption{Isophotal area versus luminosity of LABs at $z=3.1$, above a
surface-brightness threshold of $2.2 \times 10^{-18}$ erg s$^{-1}$ cm$^{-2}$
arcsec$^{-2}$. The simulated LABs from \CD (red filled circles) and \MN (green
open circles) are compared with the 35 observed LABs from \citet{matsuda} (blue
filled squares). The pink solid line refers to a power law of slope $0.75$,
which resembles the trend seen both in the simulations and the observations.}
\label{fig:area}
\end{figure}

\Fig{area} displays the areas and luminosities of our simulated LABs in
comparison with the 35 observed LABs. Each of the three simulated \CD galaxies
is ``observed" from three orthogonal directions, and each of the randomly
selected \MN galaxies is ``observed" from one random direction. We see that the
area at a given luminosity in \MN is systematically smaller than in \CD, but
only by a factor of two. The simulations agree with the observations to within
a factor of two, and they reproduce the general correlation between area and
luminosity. The overall trend follows the scaling relation  
\be
A \prop L^{0.75}  \, .
\label{eq:A-L}
\ee

\subsection{Linewidth}

The major contribution to the LAB line profile is the kinematic effect owing to
the distribution of line-of-sight velocities in the cold gas. It is dominated
by the coherent instreaming velocities from the background and from the
foreground, corresponding to a FWHM of order a few times the halo virial
velocity. The effect of resonant scattering of \lya should be convolved with
the kinematic effect.

\begin{figure}
\begin{center}
\includegraphics[width=8.4cm]{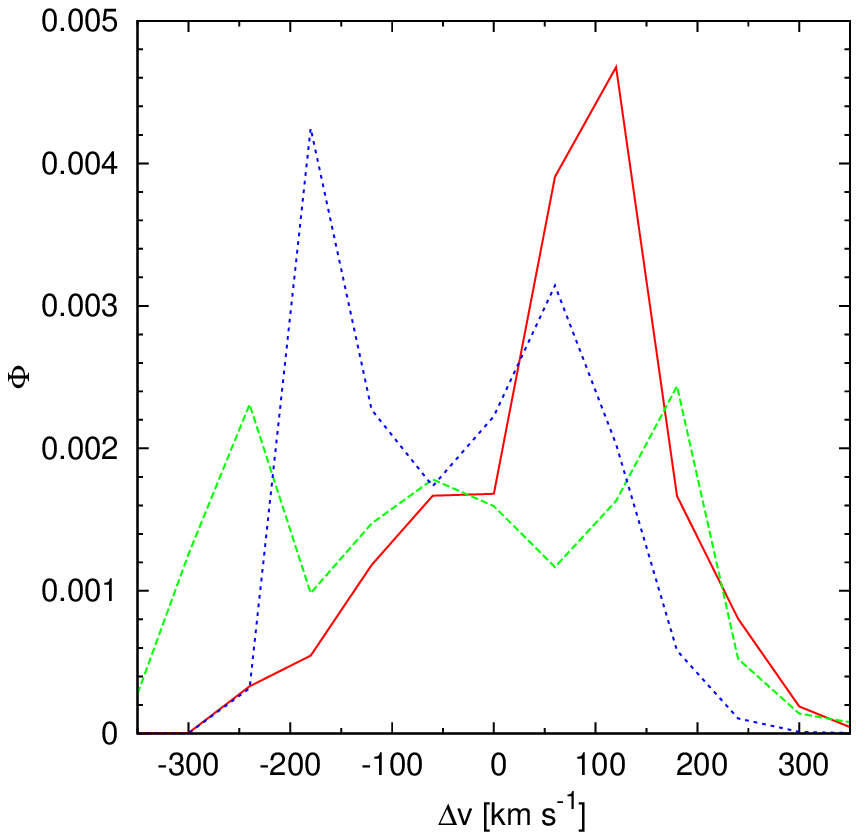}
\end{center}
\caption{The kinematic components of three characteristic \lya line profiles 
from \CD simulated galaxies, tentatively neglecting the effect of resonant
scattering. These examples have between one to three dominant peaks, reelecting
the relative orientations of the big cold streams and the line of sight.}
\label{fig:linepro}
\end{figure}

In Fig. \ref{fig:linepro} we show the kinematic components of three
characteristic line profiles from the simulated \CD galaxies. The effect of
resonant scattering is not included yet. For any desired direction of view, the
kinematic line profile is computed as the luminosity weighted distribution of
the line-of-sight velocities across the whole isophotal area. In some cases,
the line profile is dominated by one peak, corresponding to a stream that
stretches roughly along the line of sight. In other cases, the line profile is
bimodal, with a red peak and a blue peak corresponding to two opposite streams
that lie not far from the line of sight. A third type of line profile is
flatter, showing three peaks, the central of which corresponding to a stream
that is roughly perpendicular to the line of sight. We crudely estimate the
kinematic FWHM as twice the standard deviation of $\Delta v$.

Resonant scattering of the \lya photons in the stream neutral hydrogen gas
broadens the \lya emission line profile as the trapped photons diffuse into the
damping wings before finally escaping \citep{adams72, neufeld}. For scattering
and escape from a static plane parallel medium, the total line width, in units
of the Doppler parameter, is \citep{adams72,harrington}
\be
\sigma_{\rm scat} \simeq 2 [a\, \tau_{\La}\, \sqrt(3/\pi)]^{1/3} \, .
\ee
For \lya, the Doppler parameter is $b=12.89\, T_4^{1/2} \kms$, and the
radiation parameter $a$ and the optical depth $\tau_{\La}$ are given in
\se{lal} following \equ{taudust}. We obtain
\be
\sigma_{\rm scat} \simeq 436\, N_{20}^{1/3}\, T_4^{1/6} \kms \, .
\label{eq:sigma_scat}
\ee 
Thus, the line profile resulting from resonant scattering of a midplane source
is expected to consist of a red and a blue peak separated by $\sigma_{\rm
scat}$. For a very crude estimate of total line width $\Delta v$, we add in
quadrature $\sigma_{\rm scat}$ to the FWHM of the kinematic line
profile.\footnote{We note that the effect of resonant scattering as determined
for a static medium may be an overestimate because of turbulent motions.} 
 
\begin{figure}
\begin{center}
\includegraphics[width=8.4cm]{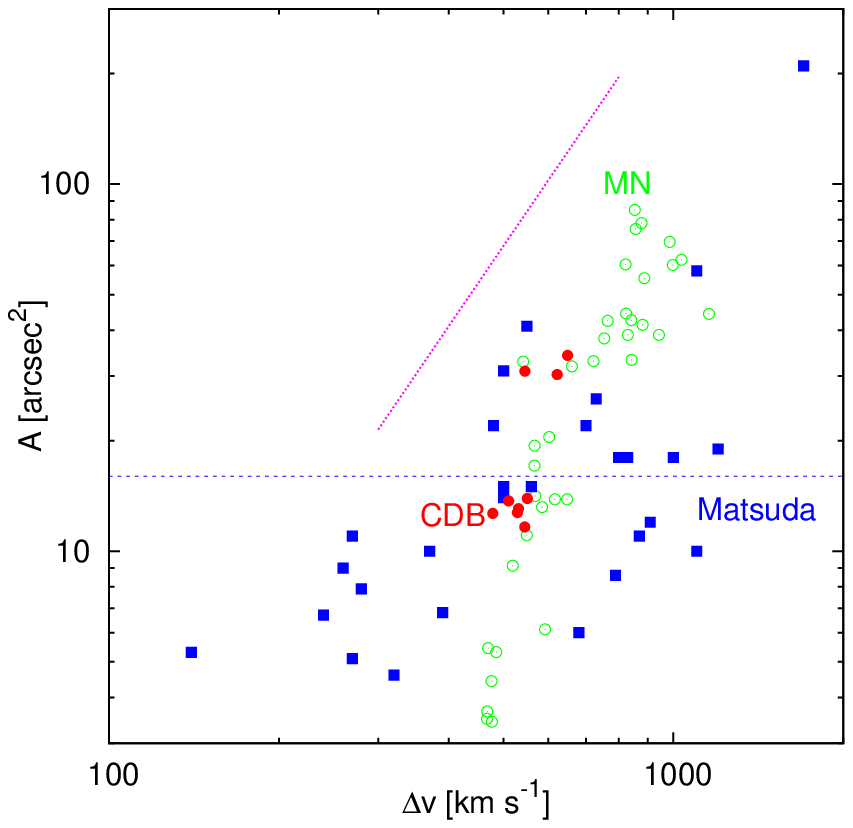}
\end{center}
\caption{Isophotal area versus linewidth (FWHM) for LABs at $z=3.1$. The
simulated LABs from \CD (red filled circles) and \MN (green open circles) are
compared with \lya emitters from \citet{matsudab} (blue filled squares). The
pink solid line refers to a power law of slope $2.25$, which crudely resembles
the general trend seen in the simulations and in the observations. The $\Delta
v$ deduced from the simulations is an underestimate and is subject to an
artificially sharp lower bound at $436\kms$.}
\label{fig:fwhm}
\end{figure}

\Fig{fwhm} displays isophotal area versus linewidth for the simulated LABs and
for observed \lya emitters \citet{matsudab}, including LABs and less extended
emitters. The simulations and observations show a similar general trend, which
very crudely follows $A \propto (\Delta V)^{2.25}$, but the data show large
scatter. A comparison to the tighter correlation seen in \fig{area} between the
area and the luminosity indicates that the scatter in \fig{fwhm} is mostly due
to scatter in $\Delta v$. This scatter is partly due to the variations in the
relative orientations of the streams and the line of sight. The spread in the
$\Delta v$ as estimated in the simulations is underestimated due to the fact
that we added in quadrature a constant value of $\sigma_{\rm scat}=436\kms$.
This is also responsible for the artificially sharp lower bound at $\Delta v =
436 \kms$, which is apparent for small values of $A$. 

\begin{figure}
\begin{center}
\includegraphics[width=8.4cm]{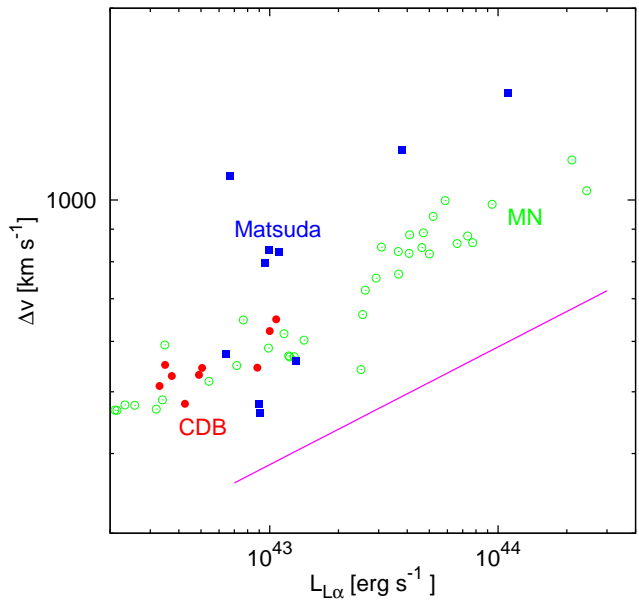}
\end{center}
\caption{\lya luminosity versus linewidth (FWHM) for LABs at $z=3.1$. The
simulated LABs from \CD (red filled circles) and \MN (green open circles) are
compared with LABs from \citet{matsuda} and \citet{matsudab} (blue filled
squares). The pink solid line refers to a power law of slope $0.18$, which
crudely resembles the general trend seen in the simulations and in the
observations, which are consistent within a factor of two.}
\label{fig:fwhmvsl}
\end{figure}

Finally, \fig{fwhmvsl} displays linewidth versus luminosity for the simulations
and a subset of the observed LABs by \citet{matsudab}. The simulations reveal a
scaling relation that can be approximated by $\Delta v \propto L^{0.18}$. The
simulations and observations agree to within a factor better than two, but the
larger scatter in the observed data confirms our suspicion that we have
underestimated the scatter in the analysis of $\Delta v$ from the simulations.

We conclude that our simplified analysis of \lya emission from simulated
galaxies qualitatively reproduces the observed correlations between the global
quantities of \lya luminosity, area and linewidth. A more detailed comparison
of theory and observed LABs, especially involving line profiles and linewidths,
should await a more accurate analysis of the simulations using radiative
transfer and including dust absorption, as well as photometric and
spectroscopic measurements of a larger sample of observed LABs.

\section{Gravitational Heating}
\label{sec:grav}

We now return to the energy source for the \lya emission in our simulations. In
addition to the gravitational heating, which was crudely approximated in
\se{feasibility}, there is photoionisation by the UV background.
Photoionisation effects by central sources such as stars and AGN were not
simulated, and they are not expected to be substantial because of the shielding
implied by the radial orientation of the cold streams. Before we address
gravitational heating again, we should verify the contribution of the UV
background. 

A simple estimate is as follows. For an isotropic background with mean photon
intensity $4 \pi J^*$ and mean photon energy $e$, the rate at which energy is
absorbed across a spherical surface of radius $R$ is $\pi J^*\, e\, 4 \pi R^2$.
If we adopt for the metagalactic field at $z=3$ a flux $\pi J^* \sim 5.5 \times
10^4 {\rm photons}\, {\rm s}^{-1}\, {\rm cm}^{-2}$ and $e \sim 20 {\rm eV}$, we
obtain a total UV heating rate into the virial radius $\Rv \sim 100 \kpc$ of $2
\times 10^{42} \ergs$. Even if all of the UV energy goes into heating the cold
streams, the UV heating is small compared to the total \lya luminosity of $\sim
10^{43} \ergs$ and to the gravitational heating rate as estimated in
\se{feasibility}. However, one should keep in mind that the {\MN} simulation
may overpredict the energy coming from the UV background when assuming that the
gas is optically thin and the UV background is uniform.

\begin{figure}
\begin{center}
\includegraphics[width=8.4cm]{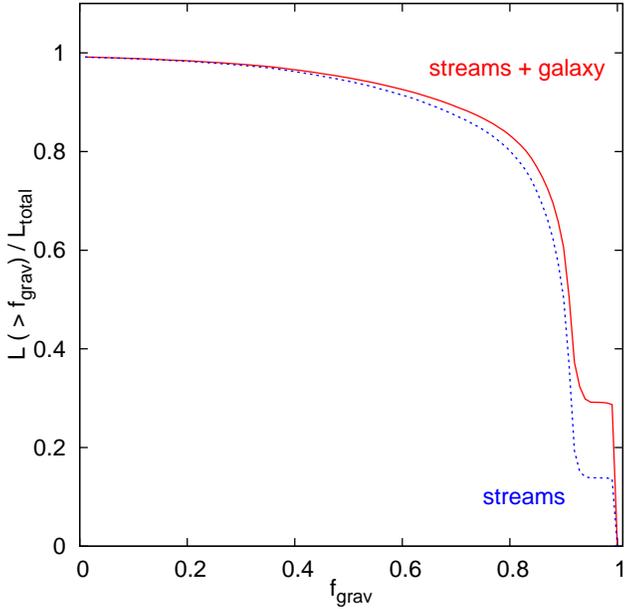}
\end{center}
\caption{The role of gravitational heating versus photoionisation. Shown is the
cumulative, luminosity-weighted distribution of the fraction of the energy that
is provided by gravitational heating, $\fg$, in the three {\CD} galaxies. The
curves refer to the streams at $r=(0.2-1)\Rv$ (dashed blue) and to the whole
halo including the inner galaxy (solid red). The area under the curve implies
that for the total {\CD} halo within $R_{\rm v}$ 86.8\% of the energy input is
gravitational, and only 13.2\% is due to the UV background. The numbers are
84.2\% and 15.8\% respectively for the {\CD} halo between 0.2 and 1.0
$R_{\rm v}$.}
\label{fig:uvgrav}
\end{figure}

To evaluate the relative contributions of gravitational heating and UV
background in our actual simulation, we read from each grid cell of the
simulation snapshot the instantaneous input rate of energy per unit volume by
the UV flux, $\epsilon_{\rm UV}$, and subtract it from the \lya emissivity
computed in \equ{eps}, to obtain the relative contribution of gravitational
heating in that cell,
\be
\fg \equiv \frac{\epsilon_{\La}-\epsilon_{\rm UV}}{\epsilon_{\La}} \ .
\ee
\Fig{uvgrav} shows the cumulative luminosity-weighted distribution of $\fg$ for
the three {\CD} galaxies at $z=2.3$. In this plot $\fg=0$ refers to cells where
all the \lya luminosity is driven by the UV background and $\fg=1$ corresponds
to gravitational heating only. The area under the curve is the fraction of the
total energy provided by gravity. In the \CD haloes this fraction is 86.8\%,
with only 13.2\% due to the UV background for the total halo within $R_{\rm
v}$. It is 84.2\% and 15.8\% respectively for the halo between 0.2 and 1.0
$R_{\rm v}$. In more detail, if we focus, for example, on the cells of gas with
the highest values of $\fg$ that contribute 0.8 of the total luminosity, we
read from the figure that they all have $\fg> 0.8$. This means that in the gas
that is responsible for 80\% of the \lya luminosity, more than 80\% of the
energy is gravitational and less than 20\% is due to the UV background. The
fraction of the total energy provided by gravity in the \MN haloes is 91.4\%,
with only 8.6\% due to the UV background for the total halo within $R_{\rm v}$.
It is 79.5\% and 20.5\% respectively for the halo between 0.2 and 1.0
$R_{\rm v}$. We conclude that most of the \lya emission from our simulated
galaxies is indeed driven by gravitational heating.

The total luminosity as predicted by the simplified gravitational-heating toy
model can be compared to the total \lya luminosity that is actually produced in
the simulations. We focus on the halo streams in the radius range $r=(0.2-1.0)
\Rv$ in haloes of $\Mv = 10^{12}\msun$ at $z = 3$. The toy-model estimate,
based on \equ{eheat} and \fig{gravpot}, is $E_{\rm heat} \sim 10^{43}\ergs$. A
straightforward integral of the emissivity in the simulated galaxies over the
same volume in the halo yields a comparable value (which can be read as $L/\fa$
from \fig{lumiz246}). This implies that the toy-model parameter $\fc$ should be
of order unity, indicating that most of the gravitational energy is deposited
in the cold streams.

We now return to \fig{gravpot}, where we showed the gravitational heating
toy-model prediction for the cumulative 3D luminosity profile. Shown in
comparison is the corresponding \lya luminosity profile as derived from the
actual simulated \CD galaxies, averaged over the three galaxies. The similarity
of the two profiles is remarkable. This implies, again, that the \lya emission
is indeed powered by gravitational heating, where the potential energy of the
instreaming cold gas is radiated as \lya.

The cumulative luminosity profile can be fitted by a power law,
\be
L(<r) \propto r^{\alpha} \, ,
\label{fig:Lprofile}
\ee
corresponding to a 3D luminosity density profile $\ell(r)\propto r^{\alpha-3}$.
In the range $r=(0.1-0.4)\Rv$, we find for the simulated luminosity $\alpha
\simeq 0.8$, namely $\ell \propto r^{-2.2}$. This is consistent with the
average surface-brightness profile $I(r) \sim r^{-1.2}$ measured in
\se{profile}. The toy model predicts a similar power index over the same radius
range. We also note that the density profile of cold gas follows a similar
power law.

\section{Discussion and Conclusion}
\label{sec:con}

Hydrodynamical cosmological simulations robustly demonstrate that massive
galaxies at high redshifts were fed by cold gas streams, inflowing into
dark-matter haloes at high rates along the cosmic web \citep{keresa, db06,
nature}. In this paper we have shown that these streams should be observable as
luminous \lya sources, with elongated irregular structures stretching for
distances of over 100 kpc. The release of gravitational potential energy by the
instreaming gas as it falls into the halo potential well is the origin of the
\lya luminosity, which ranges from $10^{42}$ to $10^{44}\ergs$ for haloes in
the mass range $10^{11}-10^{13}\msun$ at $z \sim 3$. The predicted \lya emission
morphologies and luminosities make such streams likely candidates for the
sources of observed high redshift \lya blobs. Most of the gas in the cold
streams is at temperatures in the range $(1-5)\times 10^4$K, for which the \lya
emissivity is maximised. The hydrogen gas densities in the streams are in the
range $0.01-0.1 \cmpc$, and are higher in clumps that flow with the streams,
leading to the high surface brightnesses.

Using state-of-the-art cosmological AMR simulations with 70 pc resolution and
cooling to below $10^4$K, and applying a straightforward analysis of \lya
emissivity due to electron impact excitation, we produced maps of \lya emission
from simulated massive galaxies at $z \sim 3$. We computed the average \lya
luminosity per given halo mass and the predicted luminosity function of these
extended \lya sources, with an uncertainty of $\pm 0.4$ dex. The properties of
the individual images resemble those of the observed LABs in terms of
morphology and kinematics. The predicted luminosity function is close to the
observed LAB luminosity function. The simulated LABs qualitatively reproduce
the observed correlations between the global LAB properties of \lya luminosity,
isophotal area and linewidth.

The LAB properties can be understood using a very simple toy model that
accounts for gravitational heating of the inflowing gas. In comparison, the
more elaborate toy model of DL09 predicts a steeper power-law relation between
luminosity and halo mass (their Fig.~3, with $f_{\rm g}=0.2$). In the mass
range $10^{12} - 10^{13}\msun$, DL09 predict luminosities in the range $1.5
\times 10^{42} - 2.5\times 10^{44} \ergs$. A comparison to our \fig{lumiz246}
indicates that they underestimate the luminosity at $\Mv \sim 10^{12}\msun$ by
a factor of a few. The DL09 toy model matches the observed luminosity function
only after the predictions are boosted, trying to account for an overdense
region of the universe. The association of the predicted LABs with massive
haloes places them preferentially in overdense environments, as observed
\citep{steidel, matsuda, matsudab}, but this is already 
accounted for by the number density of haloes in a fair sample.

The association of observed LABs with star-forming LBGs \citep{hayashino},
sub-millimetre galaxies \citep{chapman, geacha} or active galactic nuclei
\citep{geachc, basu} is not surprising since star formation and AGN activity
are triggered by the same process of streaming of cold gas into massive haloes.
\lya emission can be driven by any of these sources of energy. For example, in
photoionised gas in star-forming regions the \lya luminosity is $L_{\rm
L\alpha} = 1.5 \times 10^{42}\, {\rm SFR}\, {\rm erg}\, {\rm s}^{-1}$ where SFR
is the star-formation rate (in $M_\odot$ yr$^{-1}$) for continuous star
formation for a Kroupa initial mass function \citep{shp}. Thus the typical LAB
luminosity of 10$^{43}$ erg s$^{-1}$ would require SFR $\sim 10$ M$_\odot$
yr$^{-1}$. However, most of this radiation is likely absorbed by dust, and
confined to the central galaxy.

A limitation of our AMR simulations arises from the artificial pressure floor
imposed in order to properly resolve the Jeans mass. This may have an effect on
the temperature and density of the cold gas in the streams, with potential
implications on the computed \lya emission. Still, the AMR code is the best
available tool for recovering the stream properties. With 70 pc resolution and
proper cooling below $10^4$K, the {\CD} simulations provide the most reliable
description of the cold streams so far. The rather small correction that we had
to apply to the luminosities extracted from the {\MN} simulation indicates that
the \MN galaxies can be used to recover the mass and redshift dependence of the
global stream properties and the scaling relations between them.

Another source of uncertainty has to do with the simplified way the ionisation
by the UV background is handled in the simulations, and with the
post-processing calculation of the ionisation state of the cold gas. In the \CD
simulations, the centres of the streams, where the gas density is higher than
$0.1 \cmmc$, were assumed to be self-shielded against the UV background, in
agreement with our analytic estimates in \se{lal}. The ionisation state of the
gas, which is a key ingredient in evaluating the \lya emissivity, was computed
in post-processing assuming CIE. Cooling rates were computed for the given gas
density, temperature, metallicity, and UV background based on \textsc{Cloudy}
(\CD) or assuming ionisation equilibrium for H and He, including both
collisional- and photo-ionisation (\MN). Photoionisation is neglected since the
streams are sufficiently thick to be self-shielded. An alternative calculation
using \textsc{Cloudy} yielded lower fractions of neutral hydrogen at $n \leq
0.01 \cmmc$, and total luminosities that are typically three times smaller than
obtained using CIE. This calculation assumed that each cell, with a given
hydrogen density $n$, is in the middle of a uniform slab of thickness 1 kpc.
Based on our estimates of self-shielding, \equ{shielding}, we adopt the CIE
results as the more reliable approximation.

In our computations so far we did not consider the radiative transfer of \lya
photons through the streams, or through the intervening intergalactic medium.
We assume that most of the radiation emitted at $z\sim 3$ will reach the
observer at $z=0$ without undergoing significant attenuation. Resonant line
scattering in the optically thick streams will tend to spread out the \lya
emission region, while the repeated Doppler shifts broaden the line profiles,
and the increased path length of the random walk amplifies the absorption by
dust.  Such effects are expected to modify the images and line profiles, but to
have only a small effect on the total \lya luminosity. An analysis of \lya
emission from our simulated galaxies including the effects of radiative
transfer and dust absorption and a more accurate treatment of photoionisation
from stars will be presented in a forthcoming paper (Kasen et al, in
preparation).

Other hydrogen emission lines, such as L$\beta$ or H$\alpha$, are expected to
be two orders of magnitude less luminous than \lya in our model \citep{baldwin,
miller2}. The column densities of $N_{\rm HI}\sim 10^{20}\cmms$ in the cold
streams should also be detectable as \lya absorption in quasar spectra
\citep{prochaska99, wolfe}. Considering all the galaxies that are intersected
by a line of sight to a background quasar with an impact parameter $<\Rv$, we
crudely estimate from the simulations that absorption by $N_{20}>1$ hydrogen
should be detected in $\sim 20\%$ of the galaxies. Alternatively, \lya photons
that are emitted form the central galaxy should be absorbed in the radial
streams feeding the same galaxy at $N_{20}>10$, but only in $\sim 5\%$ of the
galaxies. The streams could also be detectable as low-ionisation metal
absorbers \citep[e.g.][]{gnat} as long as the metallicity in the streams is
greater than $\sim 0.01$ solar (paper in preparation).

Our results support the idea that the observed LABs are direct detections of
the cold steams that drive the evolution of massive galaxies at high redshifts.
Even though the observed LABs are sometimes associated with central sources
that are energetic enough to power the observed \lya emission, such as
starbursts and AGNs, these central sources are very different from each other
in the different galaxies, and in many LABs they are absent altogether
\citep{yang, geachb}. The gravitational heating associated with the inflowing
cold streams is a natural mechanism for driving the extended \lya cooling
radiation observed as LABs, and this extended \lya emission is inevitable in
most high-redshift galaxies.

\section*{Acknowledgements}
The computer simulations were performed at NERSC, LBNL, and at the Barcelona
Centro Nacional de Supercomputaci{\'o}n as part of the Horizon collaboration. 
We thank Yuchi Matsuda, Kim Nilson and Masami Ouchi for sharing observational
data with us. We acknowledge stimulating discussions with Nicolas Bouch{\'e},
Michele Fumagalli, Orly Gnat, Daniel Kasen, Anatoly Klypin, Kamson Lai, Piero
Madau, Ari Maller, Mark Mozena, Eyal Neistein, Hagai Netzer, and Jason X.
Prochaska. This research has been partly supported by an ISF grant, by GIF
I-895-207.7/2005, by a France-Israel Teamwork in Sciences, by the Einstein
Center at HU, by NASA ATP NAG5-8218 at UCSC. We thank the DFG for support via
German-Israeli Project Cooperation grant STE1869/1-1.GE625/15-1. Tobias Goerdt
is a Minerva fellow and Daniel Ceverino is a Golda Meir fellow.

\bibliographystyle{mn2e}
\bibliography{lya22.bbl}

\begin{thebibliography}{}

\bibitem[\protect\citeauthoryear{Adams}{1972}]{adams72}
Adams T. F, 1972, ApJ, 174, 439

\bibitem[\protect\citeauthoryear{Adams}{1975}]{adams75}
Adams T. F, 1975, ApJ, 201, 439

\bibitem[\protect\citeauthoryear{Baldwin}{1977}]{baldwin}
Baldwin J. A, 1977, MNRAS, 178, 67

\bibitem[\protect\citeauthoryear{Basu-Zych \& Scharf}{2004}]{basu}
Basu-Zych A, Scharf C, 2004, ApJ, 615, 85

\bibitem[\protect\citeauthoryear{Birnboim, Dekel \& Neistein}{2007}]{bdn07}
Birnboim Y, Dekel A, Neistein E, 2007, MNRAS, 380, 339

\bibitem[\protect\citeauthoryear{Birnboim \& Dekel}{2003}]{bd03}
Birnboim Y, Dekel A, 2003, MNRAS, 345, 349

\bibitem[\protect\citeauthoryear{Bullock et al.}{2001}]{bullock01}
Bullock J. et al, 2001, MNRAS, 321, 559

\bibitem[\protect\citeauthoryear{Callaway, Unnikrishnan \& Oza}{1987}]
{callaway87}
Callaway J, Unnikrishnan K, Oza D. H, 1987, PhRvA, 36, 2576

\bibitem[\protect\citeauthoryear{Ceverino-Rodriguez}{2008}]{ceverino}
Ceverino-Rodriguez D, 2008, Ph.D.~Thesis

\bibitem[\protect\citeauthoryear{Ceverino, Dekel \& Bournaud}{2009}]{cd}
Ceverino D, Dekel A, Bournaud F, 2009, arXiv:0907.3271 [astro-ph.CO]

\bibitem[\protect\citeauthoryear{Ceverino \& Klypin}{2009}]{cak}
Ceverino D, Klypin A. A, 2009, ApJ, 695, 292

\bibitem[\protect\citeauthoryear{Chapman et al.}{2001}]{chapman}
Chapman S. C, Lewis G. F, Scott D, Richards E, Borys C, Steidel C. C,
Adelberger K. L, Shapley A. E, ApJ, 548, 17

\bibitem[\protect\citeauthoryear{Dalla Vecchia \& Schaye}{2008}]{dvs}
Dalla Vecchia C, Schaye J, 2008, MNRAS, 387, 1431

\bibitem[\protect\citeauthoryear{Dekel \& Birnboim}{2006}]{db06}
Dekel A, Birnboin Y, 2006, MNRAS, 368, 2

\bibitem[\protect\citeauthoryear{Dekel \& Birnboim}{2008}]{db08}
Dekel A, Birnboin Y, 2008, MNRAS, 383, 119

\bibitem[\protect\citeauthoryear{Dekel et al.}{2009}]{nature}
Dekel A, Birnboin Y, Engel G. et al, 2009, Nature, 457, 451

\bibitem[\protect\citeauthoryear{Dekel, Sari \& Ceverino}{2009}]{dsc}
Dekel A, Sari R, Ceverino D, 2009, 2009, ApJ, 703, 785

\bibitem[\protect\citeauthoryear{Dijkstra \& Loeb}{2009}]{dijkstra}
Dijkstra M, Loeb A, 2009, MNRAS, 400, 1109, DL09

\bibitem[\protect\citeauthoryear{Draine}{1984}]{draine84}
Draine B, Lee H. M, 1984, ApJ, 285, 89

\bibitem[\protect\citeauthoryear{Draine}{2003}]{draine03}
Draine B, 2003, ARA\&A, 41, 241

\bibitem[\protect\citeauthoryear{Dubois \& Teyssier}{2008}]{dubois}
Dubois Y, Teyssier R, 2008, A\&A, 477, 79

\bibitem[\protect\citeauthoryear{Elmegreen et al.}{2007}]{elmegreen}
Elmegreen D. M, Elmegreen B. G, Ferguson T, Mullan B, 2007, ApJ, 663, 734

\bibitem[\protect\citeauthoryear{Fardal et al.}{2001}]{fardal}
Fardal M. A, Katz N. G, Jeffrey P, Hernquist L, Weinberg D. H, Dav{\' e} R,
2001, ApJ, 562, 605

\bibitem[\protect\citeauthoryear{Faucher-Giguere et al.}{2008b}]{faucherb}
Faucher-Giguere C, Prochaska J. X, Lidz A, Hernquist L, Zaldarriaga A,
2008b, ApJ, 681, 831

\bibitem[\protect\citeauthoryear{Ferland et al.}{1998}]{ferland}
Ferland G. J, Korista K. T, Verner D. A, Ferguson J. W, Kingdon J. B, Verner E.
M, 1998, PASP, 110, 761

\bibitem[\protect\citeauthoryear{F\"orster Schreiber et al.}{2006}]{foerster}
F\"orster Schreiber N. M. et al, 2006, ApJ, 645, 1062

\bibitem[\protect\citeauthoryear{F\"orster Schreiber et al.}{2009}]{foerster2}
F\"orster Schreiber N. M. et al, 2009, ApJ, 706, 1364

\bibitem[\protect\citeauthoryear{Furlanetto et al.}{2005}]{furlanetto}
Furlanetto S. R, Schaye J, Springel V, Hernquist L, 2005, ApJ, 622, 7

\bibitem[\protect\citeauthoryear{Furlanetto et al.}{2003}]{furlanetto2}
Furlanetto S. R, Schaye J, Springel V, Hernquist L, 2003, ApJ, 599, 1

\bibitem[\protect\citeauthoryear{Geach et al.}{2005}]{geacha}
Geach J. E. et al, 2005, MNRAS, 363, 1398

\bibitem[\protect\citeauthoryear{Geach et al.}{2007}]{geachb}
Geach J. E, Smail I, Chapman S. C, Alexander D. M, Blain A. W, Stott J. P,
Ivison R, 2007, ApJ, 655, 9

\bibitem[\protect\citeauthoryear{Geach et al.}{2009}]{geachc}
Geach J. E. et al, 2009, ApJ, 700, 1

\bibitem[\protect\citeauthoryear{Genel et al.}{2008}]{genel}
Genel S. et al, 2008, ApJ, 688, 789

\bibitem[\protect\citeauthoryear{Genzel et al.}{2006}]{genzel06}
Genzel R. et al, 2006, Nature, 442, 786

\bibitem[\protect\citeauthoryear{Genzel et al.}{2008}]{genzel08}
Genzel R. et al, 2008, ApJ, 687, 59

\bibitem[\protect\citeauthoryear{Gnat \& Sternberg}{2007}]{gnat}
Gnat O, Sternberg A, 2007, ApJS, 168, 213

\bibitem[\protect\citeauthoryear{Grevesse \& Sauval}{1998}]{grevesse}
Grevesse N, Sauval A. J, 1998, SSRv, 85, 161

\bibitem[\protect\citeauthoryear{Haardt \& Madau}{1996}]{haardt}
Haardt F, Madau P, 1996, ApJ, 461, 20

\bibitem[\protect\citeauthoryear{Haiman, Spaans \& Quataert}{2000}]{haiman}
Haiman Z, Spaans M, Quateart E, 2000, ApJ, 537, 5

\bibitem[\protect\citeauthoryear{Haiman \& Rees}{2001}]{haimanb}
Haiman Z, Rees M. J, 2001, ApJ, 556, 87

\bibitem[\protect\citeauthoryear{Harrington}{1973}]{harrington}
Harrington J. P, 1973, MNRAS, 162, 43

\bibitem[\protect\citeauthoryear{Hayashino et al.}{2004}]{hayashino}
Hayashino T. et al, 2004, AJ, 128, 2073

\bibitem[\protect\citeauthoryear{Hui \& Gnedin}{1997}]{hui}
Hui L, Gnedin N. Y, 1997, MNRAS, 292, 27

\bibitem[\protect\citeauthoryear{Jimenez \& Haiman}{2006}]{jimenez}
Jimenez R, Haiman Z, 2006, Nature, 440, 501

\bibitem[\protect\citeauthoryear{Johansson, Naab \& Ostriker}{2009}]{peter}
Johansson P. H, Naab T, Ostriker J. P, 2009, ApJ, 697, 38

\bibitem[\protect\citeauthoryear{Katz, Hernquist \& Weinberg}{1992}]{katz}
Katz N, Hernquist L, Weinberg D. H, 1992, ApJ, 399, 109

\bibitem[\protect\citeauthoryear{Kennicutt}{1998}]{kennicutt}
Kennicutt R. C, 1998, ApJ, 498, 541

\bibitem[\protect\citeauthoryear{Keres et al.}{2005}]{keresa}
Keres D, Katz N, Weinberg D. H, Dav{\'e} R, 2005, MNRAS, 363, 2

\bibitem[\protect\citeauthoryear{Keres et al.}{2009}]{keresb}
Keres D, Katz N, Fardal M, Dav{\'e} R, Weinberg D. 2009, MNRAS, 395, 160

\bibitem[\protect\citeauthoryear{Khochfar \& Ostriker}{2008}]{ko08}
Khochfar S, Ostriker J. P, 2008, ApJ, 680, 54

\bibitem[\protect\citeauthoryear{Komatsu et al.}{2009}]{WMAP5}
Komatsu E. et al, 2009, ApJS, 180, 330

\bibitem[\protect\citeauthoryear{Kravtsov}{2003}]{andrey}
Kravtsov A. V, 2003, ApJ, 590, 1

\bibitem[\protect\citeauthoryear{Kravtsov, Klypin \& Khokhlov}{1997}]{kkk}
Kravtsov A. V, Klypin A. A, Khokhlov A. M, 1997, ApJS, 111, 73

\bibitem[\protect\citeauthoryear{Matsuda et al.}{2004}]{matsuda}
Matsuda Y. et al, 2004, AJ, 128, 569

\bibitem[\protect\citeauthoryear{Matsuda et al.}{2006}]{matsudab}
Matsuda Y, Yamada T, Hayashino T, Yamauchi R, Nakamura Y, 2006, ApJ, 640, 123

\bibitem[\protect\citeauthoryear{Miller \& Scalo}{1979}]{miller}
Miller G. E, Scalo J. M, 1979, ApJS, 41, 513

\bibitem[\protect\citeauthoryear{Miller}{1974}]{miller2}
Miller J. S, 1974, ARA\&A, 12, 331

\bibitem[\protect\citeauthoryear{Mori et al.}{2004}]{mori}
Mori M, Umemura M, Ferrara A, 2004, ApJ, 613, 97

\bibitem[\protect\citeauthoryear{Navarro, Frenk \& White}{1997}]{nfw}
Navarro J. F, Frenk C. S, White S. D. M, 1997, ApJ, 490, 493

\bibitem[\protect\citeauthoryear{Neistein, van den Bosch \& Dekel}{2006}]
{neistein06}
Neistein E, van den Bosch F, Dekel A, 2006, MNRAS, 372, 933

\bibitem[\protect\citeauthoryear{Neistein \& Dekel}{2008}]{neistein07}
Neistein E, Dekel A, 2008, MNRAS, 383, 615

\bibitem[\protect\citeauthoryear{Neufeld}{1990}]{neufeld}
Neufeld D. A, 1990, ApJ, 350, 216

\bibitem[\protect\citeauthoryear{Nilson et al.}{2006}]{kim}
Nilsson K. K, Fynbo J. P. U, M{\o}ller P, Sommer-Larsen J, Ledoux C, 2006,
A\&A, 452, 23

\bibitem[\protect\citeauthoryear{Ocvirk, Pichon \& Teyssier}{2008}]{ocvirk}
Ocvirk P, Pichon C, Teyssier R, 2008, MNRAS, 390, 1326

\bibitem[\protect\citeauthoryear{Ouchi et al.}{2009}]{ouchi}
Ouchi M, Ono Y, Egami E. et al, 2009, ApJ, 696, 1164

\bibitem[\protect\citeauthoryear{Prochaska}{1999}]{prochaska99}
Prochaska J. X, 1999, ApJL, 511, L71

\bibitem[\protect\citeauthoryear{Rees}{1989}]{rees}
Rees M. J, 1989, MNRAS, 239, 1

\bibitem[\protect\citeauthoryear{Robertson \& Kravtsov}{2008}]{rk}
Robertson B. E, Kravtsov A. V, 2008, ApJ, 680, 1083

\bibitem[\protect\citeauthoryear{Saito et al.}{2006}]{saito}
Saito T, Shimasaku K, Okamura S, Ouchi M, Akiyama M, Yoshida M, 2006, ApJ, 648,
54

\bibitem[\protect\citeauthoryear{Scarlata et al.}{2009}]{scarlata}
Scarlata C. et al, 2009, ApJ, 706, 1241

\bibitem[\protect\citeauthoryear{Schaye \& Dalla Vecchia}{1999}]{joop}
Schaye J, Dalla Vecchia C, 2008, MNRAS, 383, 1210

\bibitem[\protect\citeauthoryear{Scharf et al.}{2003}]{scharf}
Scharf C, Smail I, Ivison R, Bower R, van Breugel W, Roland M, 2003, ApJ,
596, 105

\bibitem[\protect\citeauthoryear{Sheth \& Tormen}{1999}]{st}
Sheth R. K, Tormen G, 1999, MNRAS, 308, 119

\bibitem[\protect\citeauthoryear{Smith \& Jarvis}{1999}]{sj}
Smith D. J. B, Jarvis M, 2007, MNRAS, 378, 49

\bibitem[\protect\citeauthoryear{Springel \& Hernquist}{1999}]{sh}
Springel V, Hernquist L, 2003, MNRAS, 339, 289

\bibitem[\protect\citeauthoryear{Stark et al.}{2008}]{stark08}
Stark D. P, Swinbank A. M, Ellis R. S, Dye S, Smail I. R, Richard J, 2008,
Nature, 455, 775

\bibitem[\protect\citeauthoryear{Steidel et al.}{1998}]{steidelb}
Steidel C. C, Adelberger K. L, Dickinson M, Giavalisco M, Pettini M,
Kellogg M, 2000, ApJ, 532, 170

\bibitem[\protect\citeauthoryear{Steidel et al.}{2000}]{steidel}
Steidel C. C, Adelberger K. L, Shapley A. E, Pettini M, Dickinson M, Giavalisco
M, 2000, ApJ, 532, 170

\bibitem[\protect\citeauthoryear{Sternberg, Hoffmann \& Pauldrach}{2003}]{shp}
Sternberg A, Hoffmann T. L, Pauldrach A. W. A, 2003, ApJ, 599, 1333

\bibitem[\protect\citeauthoryear{Teyssier}{2002}]{teyssier}
Teyssier R, 2002, A\&A, 385, 337

\bibitem[\protect\citeauthoryear{Truelove et al.}{1997}]{truelove}
Truelove J. K, Klein R. I, McKee C. F, Holliman J. H, Howell L. H, Greenough J.
A, 1997, ApJ, 489, 179

\bibitem[\protect\citeauthoryear{Wolfe, Gawiser \& Prochaska}{2005}]{wolfe}
Wolfe A. M, Gawiser E, Prochaska J. X, 2005 ARA\&A, 43, 861

\bibitem[\protect\citeauthoryear{Woosley \& Weaver}{1995}]{woosley}
Woosley S. E, Weaver T. A, 1995, ApJS, 101, 181

\bibitem[\protect\citeauthoryear{Yang et al.}{2008}]{yang}
Yang Y, Zabludoff A, Tremonti C, Eisenstein D, Dav{\' e} R, 2009,
ApJ, 693, 1579

\bibitem[\protect\citeauthoryear{Yepes et al.}{1997}]{yepes}
Yepes G, Kates R, Khokhlov A, Klypin A, 1997, MNRAS, 284, 235

\end{thebibliography}

\label{lastpage}
\end{document}